\pdfoutput=1

\documentclass[11pt, reqno,preprint]{article}
\pdfoutput=1
\usepackage{jheppub}
\usepackage{epsfig}
\usepackage{amssymb}
\usepackage{amsmath}
\usepackage{mathrsfs}
\usepackage{hyperref}
\usepackage{multirow}
\usepackage{bbold}
\newcommand{\eq}{\begin{equation}}
\newcommand{\eqe}{\end{equation}}
\newcommand{\eqa}{\begin{eqnarray}}
\newcommand{\eqae}{\end{eqnarray}}

\def\be{\begin{equation}}
\def\ee{\end{equation}}
\def\ba{\begin{eqnarray}}
\def\ea{\end{eqnarray}}

\def\Li{\textrm{Li}}
\def\l{\langle}
\def\r{\rangle}
\def\nl{\nonumber\\}
\def\idmatrix{\mathbb{1}}
\def\LL{\mathcal{L}}
\def\Tr{\mbox{Tr}}
\def\Li{\textrm{Li}}
\def\IR{\textrm{IR}}
\def\sgnc{\textrm{sgn}_c}
\title{The two-loop six-point amplitude in ABJM theory}
\author[a,c]{S. Caron-Huot,}
\author[a,b,d]{Yu-tin Huang}
\affiliation[a]{School of Natural Sciences, Institute for Advanced
Study, Princeton, NJ 08540, USA} \affiliation[b]{Department of Physics and Astronomy, UCLA, Los Angeles, CA
90095-1547, USA}
\affiliation[c]{Niels Bohr International Academy and Discovery Center,
The Niels Bohr Institute,
Blegdamsvej 17, DK-2100 Copenhagen, Denmark}
\affiliation[d]{Michigan Center for Theoretical Physics, Randall Laboratory of Physics, 
University of Michigan, Ann Arbor, MI 48109, USA}

\emailAdd{schuot@ias.edu, yhuang@physics.ucla.edu}
\abstract{In this paper we present the first analytic computation of the six-point two-loop amplitude of ABJM theory. We show that the two-loop amplitude consist of corrections proportional to two distinct local Yangian invariants which can be identified as the tree- and the one-loop amplitude respectively. The two-loop correction proportional to the tree-amplitude is identical to the one-loop BDS result of $\mathcal{N}=4$ SYM plus an additional remainder function, while the correction proportional to the one-loop amplitude is finite. Both the remainder and the finite correction are dual conformal invariant, which 
implies that the two-loop dual conformal anomaly equation for ABJM is again identical to that of one-loop $\mathcal{N}=4$ super Yang-Mills, as was first observed at four-point. We discuss the theory on the Higgs branch, showing that its amplitudes are infrared finite, but equal, in the small mass limit, to those obtained in dimensional regularization.}
\preprint{ MCTP-12-24}
\begin{document}
\maketitle

\pagebreak
\section{Introduction}
Amidst the shadow of tremendous progress in $\mathcal{N}=4$ super Yang-Mills (SYM$_{4}$) amplitudes, three-dimensional Chern-Simons matter (CSM) theory has recently enjoyed a quiet surge of interest. This reflects an interesting dual aspect of the latter: On the one hand it is a close cousin to SYM theory in four-dimensions and thus provides a fruitful arena to apply the methods that was developed there-in. On the other, while scattering amplitudes of SYM$_{4}$ theory, both perturbative and non-perturbative, are closely related to string theory scattering amplitudes, such relations for CSM theory are either obscure or in some cases simply absent as the proper correspondence is with M-theory instead. The latter is intriguing in that it implies that certain novel properties that is shared between the scattering amplitudes of both Yang-Mills and CSM may in fact have a deeper purely field theoretical origin.

A prominent example is the $\mathcal{N}=6$ theory constructed by Aharony, Bergman, Jafferis and Maldacena (ABJM)~\cite{Aharony:2008ug}. Being dual to type IIA string theory in $AdS_{4}\times \mathbf{CP}^3$ background, it is very similar to SYM$_{4}$ in terms of providing an exact AdS/CFT pair. This similarity inspired the discovery of many common features between the two theories such as the presence of a hidden Yangian symmetry~\cite{Bargheer:2010hn} (or equivalently dual superconformal symmetry~\cite{DualConformal0, DualConformal,Huang:2010qy}) of the tree- and planar loop-amplitudes~\cite{Gang:2010gy},\footnote{At this stage it is unclear what role, if any, AdS/CFT  plays in the existence of these symmetries, as explicit attempts at proving self-T-duality~\cite{FermiT} on the string or supergravity side have encounter technical difficulties and have not been fully carried out~\cite{FermiT2}.} as well as the realization that the leading singularities of both theories are encoded by the residues of a contour integral over Grassmaniann manifolds~\cite{ArkaniHamed:2009dn,Lee:2010du}.   

In many aspects, ABJM amplitudes are simpler than its four-dimensional relative. This simplicity is already reflected in the fact that only even legged amplitudes are non-trivial~\cite{Agarwal:2008pu}. Furthermore, all one-loop amplitudes consist solely of rational functions~\cite{Talk, Bianchi:2012cq, Bargheer:2012cp, Brandhuber:2012un} (multiplied by $\pi$, in a natural normalization) while the two-loop amplitudes are of transcendentality-two-functions~\cite{Chen:2011vv, Bianchi:2011dg}. This should be compared to transcendentality -two- and four-functions for one and two-loop amplitudes respectively in SYM$_4$. As all one-loop amplitudes can be conveniently expressed in terms of a basis of massive triangle integrals, whose coefficients can be directly computed via recursion relations~\cite{Brandhuber:2012wy}, the one-loop amplitude for ABJM theory with arbitrary multiplicity is effectively ``solved".

On the other hand some properties of CSM theory, while shared with YMs theory, demand an alternative explanation other than the stringy origin currently available for the latter. Consider the color-kinematic duality~\cite{Bern:2008qj}, which leads to non-trivial amplitude relations for YMs and relates the amplitudes of the gauge theory to that of the corresponding gravity theory to all order in perturbation theory~\cite{Bern:2010ue, Bern:2010yg}. For CSM, it was shown that similar duality, although based on three-algebra~\cite{3Algebra}, is also present for the $\mathcal{N}=8$~\cite{Bargheer:2012gv} and  $\mathcal{N}<8$~\cite{Huang:2012vt} theory. While the relations implied by the duality in YMs can be traced back to monodromy relations of string amplitudes~\cite{BjerrumBohr:2009rd}, such correspondence does not exist for CSM theory since the amplitudes are not directly related to any open string amplitudes in a flat back-ground. As the color-kinematic identity allows one to obtain the amplitudes of gravity-matter theory from that of CSM theory,\footnote{Both pure Chern-Simons and gravity in three-dimensions are topological.} the fact that gravity amplitudes can be extracted from close string amplitudes, render the role of string theory even more mysterious. 
  
In this paper our main focus is the loop amplitudes of ABJM theory, in particular, the six-point one- and two-loop amplitude, in the planar (`t Hooft) limit.
It was shown in ref~\cite{Chen:2011vv, Bianchi:2011dg} that the two-loop four-point amplitude has the same functional dependence as that of the one-loop four-point SYM$_4$ amplitude. This equivalence was latter shown to persist to all orders in $\epsilon$ expansion~\cite{Allorder}. As the four-point amplitude can be uniquely determined by the dual conformal anomaly equation~\cite{Ward1, Ward2}, this results states that the anomaly equation for both ABJM and SYM$_4$, up to four-points, are identical. However, taking into account the fact that the theory is conformal, the simplicity of four-point kinematics and the transcendental requirement of the finite function, this result might be deemed accidental (although not for the all order correspondence).  At six-point, it is nontrivial that the anomaly equations should match. Furthermore, the anomaly equations fixes the result only up to homogenous terms and six-point is the first place where non-trivial invariant remainder functions might appear. Thus the six-point computation is an important piece of data to clarify these issues.

As ABJM theory consists of matter fields transforming under the bi-fundamental representation of the gauge group SU(N)$_k\times$SU(N)$_{-k}$, the amplitude has a definite parity under the exchange of Chern-Simons level $k\leftrightarrow-k$. More precisely and $L$-loop amplitude is weighted by a factor of $(4\pi/k)^{L+1}$, and hence (odd-)even-loop amplitudes are parity (even)odd. Assuming that parity is non-anomalous, in order for odd-loops to give an acceptable contribution it must compensate for its opposite parity. Since the exchanging of $k\leftrightarrow-k$ can be translated into the exchange of the gauge group, this implies that a non-vanishing odd-loop all scalar-amplitude must pickup a minus when cyclicly shifted by one-site. This is indeed the case.  

We construct the six-point integrand using leading singularity methods. Since it was shown in ref.~\cite{Gang:2010gy} that there is only one pair of leading singularity at six-points, it is straightforward to construct the integrand by choosing an integral basis consisting of integrands with uniform leading singularities. At one-loop there are two types, the one-loop box and massive triangle integrals with loop momentum dependent and independent numerators respectively. There are two distinct combinations of the leading singularity pair, the difference and the sum. The former is simply the tree amplitude while the latter is the conjugate tree-amplitude, with even and odd sites now belonging to the conjugate multiplet, denoted as $\mathcal{A}^\textrm{tree}_\textrm{6,shifted}$ since the identification of multiplets are shifted by one site.
We find that those two objects do appear in integrand.

In ref~\cite{Bianchi:2012cq, Bargheer:2012cp, Brandhuber:2012un}, the one-loop amplitude was given solely in terms of triangle integrals,
proportional to $\mathcal{A}^\textrm{tree}_\textrm{6,shifted}$.  This is valid up to order $\mathcal{O}(\epsilon)$ as the box-integrals integrate to zero at $\mathcal{O}(1)$. Having a result at one-loop that is valid to all orders in $\epsilon$ will be extremely important for the construction of the two-loop amplitude.\footnote{This was already seen for the four-point amplitude~\cite{Chen:2011vv} where the one-loop result vanishes up to $\mathcal{O}(\epsilon)$, yet it has a nontrivial box integrand. This integrand later becomes the seed of the two-loop integrand. The relevance of the $\mathcal{O}(\epsilon)$ pieces can also be understand from unitarity cuts, where such terms might combine with collinear singularity of the tree amplitudes to give non-trivial two loop contribution.}
The integrated result is proportional to a step function, which as we will see nicely captures the non-trivial topology of $3d$ massless kinematics. More precisely, massless kinematics in three-dimensions can be parameterized by points on $S^1$. For color ordered amplitudes, distinct kinematic configurations can be categorized by a ``winding number" which can be unambiguously defined. The sign function then simply distinguishes the configurations with even or odd winding number,
for a given kinematic channel.

With the one-loop integrand in hand, we construct the two-loop amplitude by simply requiring that on the maximal cut of one of the sub-loops, one obtains the full one-loop integrand. This fixes the integrand up to possible double triangles, which are further fixed by soft-collinear constraints. We compute the integrals using both dimensional reduction regularization as well as mass regularization. This mass regulator can actually be given a physical interpretation in terms of moving to the Coulomb branch of the theory and giving the scalars a vev, similar to that used for SYM$_4$~\cite{Alday:2009zm}. Interestingly, while the result for the individual integrals differ between the two schemes, they give, up to additive constant, identical results when combined into the final physical amplitude.

Using five-dimensional embedding formalism, the tree amplitude is multiplied by five-dimensional parity even integrals, while the conjugate tree-amplitude is multiplied by parity odd integrals. Introducing the cross-ratios (only two of the these are algebraically independent)
\eq
u_1=\frac{(1\cdot3)(4\cdot6)}{(1\cdot4)(3\cdot6)},\quad u_2=\frac{(2\cdot4)(5\cdot1)}{(2\cdot5)(4\cdot1)},\quad u_3=\frac{(3\cdot5)(6\cdot2)}{(3\cdot6)(5\cdot2)}\,,
\eqe
the two-loop amplitude is given as:   
\begin{equation}
\boxed{
\begin{split}
  \mathcal{A}_6^{\textrm{2-loop}}&=\left(\frac{N}{ k}\right)^2 \bigg\{\frac{\mathcal{A}_6^{\textrm{tree}}}{2}\bigg[BDS_6+R_6\bigg]
  +\frac{\mathcal{A}^\textrm{tree}_\textrm{6,shifted}} {4i}
    \bigg[\log\frac{u_2}{u_3}\log\chi_1+{\rm cyclic\times2}\bigg]  \bigg\}\,,
    \end{split}
}
\end{equation} 
where $BDS_6$ is the one-loop MHV amplitude for $\mathcal{N}=4$ SYM~\cite{Bern:1994zx, BDS}, with proper rescaling of the regulator to account for the fact this is at two-loops, and the remainder function $R_6$ is given as
$$R_6=-2\pi^2+\sum_{i=1}^3\left[\Li_2(1-u_i)+\frac12\log u_i\log u_{i{+}1} +(\arccos\sqrt{u_i})^2\right].$$
The $\chi_i$ are little-group-odd cross-ratios defined in (\ref{defChi}); we warn the reader that these variables may require
some care when analytically continuing to Minkowski kinematics.  An alternative form of the amplitude with explicit dependence on conventional invariants is given in eq.~(\ref{result1}).

The presence of the BDS result demonstrates that infrared divergence and the dual conformal anomaly equation of the two-loop ABJM theory is identical are that of one-loop SYM$_4$.  Furthermore, similar to SYM${}_4$, using the mass regulator we show how the anomaly equation can be converted into a statement of exact dual conformal symmetry in higher dimensions, with the mass playing the role of the extra dimension.

This paper is organized as follows: In section~(\ref{section2}) we lay out some basic conventions, while in section~(\ref{sec:oneloop}) we begin with the discussion of general one-loop dual conformal integrand and its integration in the embedding formalism. We then explicitly construct the one-loop six-point integrand and well as the integrated result. We end with a more detailed discussion of the properties of the one-loop amplitude in terms of the topological properties of three-dimensional kinematics. In section~(\ref{sec:integrand}), we employ leading singularity methods and soft-collinear constraints to fix the two-loop integrand. In section~(\ref{sec:higgs}) we briefly discuss two regularization schemes, dimensional reduction regularization and higgs mass regulation, with special emphasis on the latter. In section~(\ref{sec:integrals}) we will use mass regularization to explicitly compute the integrals. In section~(\ref{section7}) we combine the integrated expressions and give the complete six-point two-loop amplitude. We give a brief conclusion and discussion for future directions in section~(\ref{section8}).

\section{Conventions}\label{section2}
Since we will be interested in planar amplitudes, it is useful to define the dual coordinates
\be
 x_{i+1}-x_i = p_i.
\ee
Special interest in the $x_i$ coordinates resides in the fact that planar amplitudes in ABJM theories are invariant under the so-called
dual conformal transformations, which act as conformal transformations of the $x_i$.
To make the action of this symmetry simplest, and at the same time trivialize several operations which occur when doing loop computations,
we will systematically use the so-called embedding formalism~\cite{Embedding} (for more recent discussion, see~\cite{Siegel:2012di}).

The idea is to uplift three-dimensional $x_i$'s to (projectively identified) null five-vectors
\be
 y_i := (x_i,1,x_i^2)  \label{defa}
\ee
such that inverse propagators become the (2,3)-signature inner product
\be
 (i\cdot j):=y_i \cdot y_j:=(x_i-x_j)^2.
\ee
The group of conformal transformations SO(2,3) of three-dimensional Minkowski spacetime
is then realized linearly as the transformations of the $y_i$ which preserve this inner product.

It was shown in ref.~\cite{Gang:2010gy} that the tree-level amplitude and loop-level integrand in ABJM inverts homogeneously under dual conformal inversion: 
\eq
I\left[\mathcal{A}_n\right]=\prod_{i=1}^n\sqrt{x^2_i}\mathcal{A}_n\,.
\eqe
Due to the fact that at weak coupling the theory only has $\mathcal{N}=6$ supersymmetry, the on-shell states are organized into two different multiplets:
\eqa
\nonumber\Phi(\eta)&=&\phi^4+\eta^I\psi_I+\frac{1}{2}\epsilon_{IJK}\eta^I\eta^J\phi^K+\frac{1}{3!}\epsilon_{IJK}\eta^I\eta^J\eta^K\psi_4,\\
\bar{\Psi}(\eta)&=&\bar{\psi}^4+\eta^I\bar{\phi}_I+\frac{1}{2}\epsilon_{IJK}\eta^I\eta^J\bar{\psi}^K+\frac{1}{3!}\epsilon_{IJK}\eta^I\eta^J\eta^K\bar{\phi}_4,
\eqae
where $\eta^I$ are Grassmann variables in the fundamental of U(3)$\in$SU(4). The kinematic information are encoded in terms of SL(2,R)
spinors $\lambda^{\alpha}$, with 
\eq
s_{ij}=-\langle ij\rangle^2,\;\;\;\langle ij\rangle:=\lambda_i^\alpha\lambda_j^\beta\epsilon_{\alpha\beta}
\eqe 
where $s_{ij}=x_{i,i+2}^2$ when $j=i+1$.
Note that $x_{ij}^2$ is positive when the corresponding momentum is spacelike, while $\l ij\r^2$ is negative in that case.
For more detailed discussion of the on-shell variables $(\lambda_i^\alpha, \eta_i^I)$ see ref.~\cite{Bargheer:2010hn}. In this paper, we will use the convention where the barred multiplet sits on the odd sites. The four-point amplitude is given as~\cite{Bargheer:2010hn}:
\eq
\mathcal{A}_4(\bar{1}2\bar{3}4)=\frac{4\pi}{k}\frac{\delta^3(P)\prod_{I=1}^3\delta^2(Q^I)}{\langle12\rangle\langle23\rangle}\quad\mbox{with}\quad \delta^2(Q^I):=\sum_{1\leq i<j\leq4}\eta_i^I\langle ij\rangle\eta_j^I. \label{fourpoint}
\eqe

\section{One-loop integrand and amplitude}
\label{sec:oneloop}
Dual conformal symmetry restricts the integral basis to be constructed of SO(2,3) invariant projective integrals. At one-loop, this restricts the $n$-point amplitude to be expanded on the basis of scalar triangles with appropriate numerators, as well as scalar box integrals with numerator constructed from the five-dimensional Levi-Cevita tensor:
\eq
I_{\textit{box}}(i,j,k,l)=\int_a\frac{\epsilon(a,i,j,k,l)}{(a\cdot i)(a\cdot j)(a\cdot k)(a\cdot l)}\,.
\label{OneLoopBox}
\eqe
The integral in eq.~(\ref{OneLoopBox}) is analogous to the four-dimensional pentagon integral described in ref.~\cite{vanNeerven:1983vr}
and it integrates to zero up to order $\epsilon$ in dimension regularization~\cite{Chen:2011vv}.
To demonstrate how dimensional regularization is employed in the embedding formalism, we explicitly demonstrate this result in the following. 

We first note that eq.~(\ref{OneLoopBox}) can be rewritten using Feynman parametrization as 
\eq
I_{box}(i,j,k,l)=-\int dF ~\epsilon(i,j,k,l,\partial_{Y})\int_a\frac{\Gamma[3]}{(a\cdot Y)^3}
\label{DerivY}
\eqe
where $dF:= \prod_{i=1}^4d\alpha_i\delta(1-\sum_i\alpha_i)$ and $Y:= \alpha_1y_i+\alpha_2 y_j+\alpha_3 y_k+\alpha_4 y_l$. We now focus on the inner integral, which for the purpose of dimensional regularization, we define in $D$-dimensions:
\eq
I_0=\Gamma[3]\int_a\frac{1}{(a\cdot Y)^3}:=\Gamma[3]\int \frac{d^{D+2}a~\delta(a^2)}{i(2\pi)^D{\rm Vol}({\rm GL}(1))}\frac{1}{(a\cdot Y)^3(a\cdot I)^{D-3}}\,.
\eqe
Let us illuminate this definition of $\int_a$ by comparing it with (\ref{defa}).
First, the GL(1) symmetry can be gauge-fixed by setting the next-to-last component of $a$ to 1, at the price of a unit Jacobian.
Then the $\delta(a^2)$ factor forces the last component of $a$ to equal $x^2$, thus reducing $\int_a$ to the usual loop integration $\int \frac{d^Dx}{i(2\pi)^D}$. Finally, the factor of $i$ is removed by the Wick rotation from Minkowski to Euclidean space.

The key feature away from $D=3$ is the factor $(a\cdot I)$ where $y_I:=(\vec 0_D,0,1)$ is the infinity point.
This signals the breaking of dual conformal symmetry,
and is required to maintain the projective nature of the integrand (the GL(1) invariance) for arbitrary $D$.
This feature remains clearly visible when switching to the easily-obtained integrated expression:
\eq
I_0=\frac{\Gamma\left[3-\frac{D}{2}\right]}{(4\pi)^{\frac{D}{2}}}\frac{1}{(I\cdot Y)^{D-3}(\frac{1}{2}Y^2)^{3-\frac{D}{2}}}\,.  \label{intI0}
\eqe 
Plugging this into eq.~(\ref{DerivY}), we find that the box integral gives:
\eq
I_{box}(i,j,k,l)=\int \frac{dF}{(4\pi)^{\frac{D}{2}}}\frac{\Gamma\left[4-\frac{D}{2}\right]\epsilon(i,j,k,l,Y)}{(I\cdot Y)^{D-3}(\frac{1}{2}Y^2)^{4-\frac{D}{2}}}+(D-3)\frac{\Gamma\left[3-\frac{D}{2}\right]\epsilon(i,j,k,l,I)}{(I\cdot Y)^{D-2}(\frac{1}{2}Y^2)^{3-\frac{D}{2}}}\,.
\label{SubtleIssue}
\eqe
The first term vanishes due to the fact that $Y$ is a linear combination of the four external coordinates, while the second term is at least $\mathcal{O}(\epsilon)$ with $D=3-2\epsilon$. 

As the one loop box integral vanishes, dual conformal symmetry implies that the amplitude, up to $\mathcal{O}(\epsilon)$, can be solely expressed in terms of scalar triangles. However as discussed in the introduction, for the purpose of constructing the two-loop integrand it will be extremely useful (and actually essential) to have a  one-loop integrand valid beyond $\mathcal{O}(\epsilon)$. In the following, we will derive the full one-loop six-point integrand that includes both the scalar triangle and the tensor box integrals. We note that the form of the amplitude in terms of scalar triangles were given in~\cite{Bianchi:2012cq,Bargheer:2012cp}.

\subsection{Leading singularity and the one-loop integrand\label{LSdef}}
\begin{figure}
\begin{center}
\includegraphics[scale=0.8]{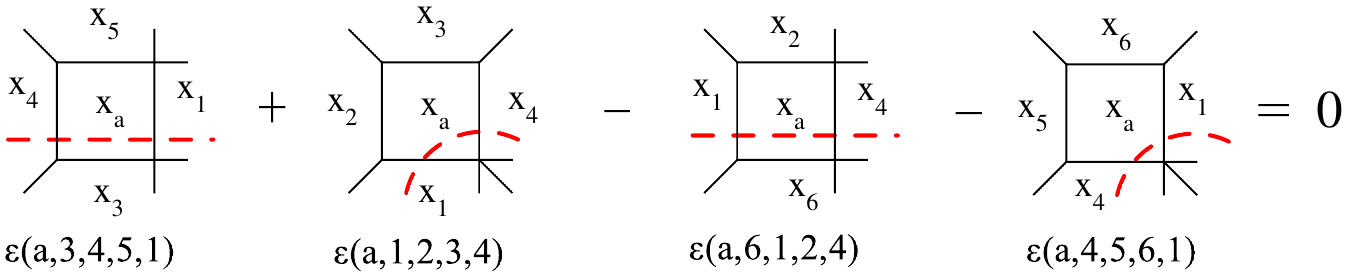}
\caption{The particular combination of tensor box integrals in eq.~(\ref{1LoopBox}) combines to give vanishing two particle cut $x^2_{a1}=x^2_{a4}=0$. This cut must vanish as the amplitude factorizes a five-point tree amplitude, which vanishes. }
\label{Coefficient}
\end{center}
\end{figure}

At six-point there are three possible box integrals, the one mass box, two-mass-easy and two-mass-hard box integrals.\footnote{Here we borrow the nomenclature of four-dimensional box integrals to denote the propagator structure.} Using the five term Schouten identity of the five-dimensional Levi-Cevita tensor one finds the following linear identity for the box integrals:
\eq
I_{box}(1,3,4,6)=I_{box}(3,4,5,6)+I_{box}(4,5,6,1)+I_{box}(1,3,5,6)+I_{box}(1,3,4,5)\,.
\eqe
Thus the two-mass-easy integral can be expressed in terms of linear combinations of the two-mass-hard and one mass integrals. We will use the later two as the basis for box integrals. The relative coefficient of the box integrals can be easily fixed by requiring that the two particle cuts which factorize the amplitude into two five-point tree amplitudes, must vanish. Cutting in the $x^2_{14}$-channel, shown in fig.~(\ref{Coefficient}), this requires four box integrals to come in the following  combination:
\eq
I_{box}(3,4,5,1)+I_{box}(1,2,3,4)-I_{box}(4,5,6,1)-I_{box}(6,1,2,4)\,.
\label{1LoopBox}
\eqe
Using the Schouten identity, one can show that this combination is actually invariant under cyclic permutation by one site up to overall sign.
This extra sign will be important as we will discuss shortly.

\begin{figure}
\begin{center}
\includegraphics[scale=0.65]{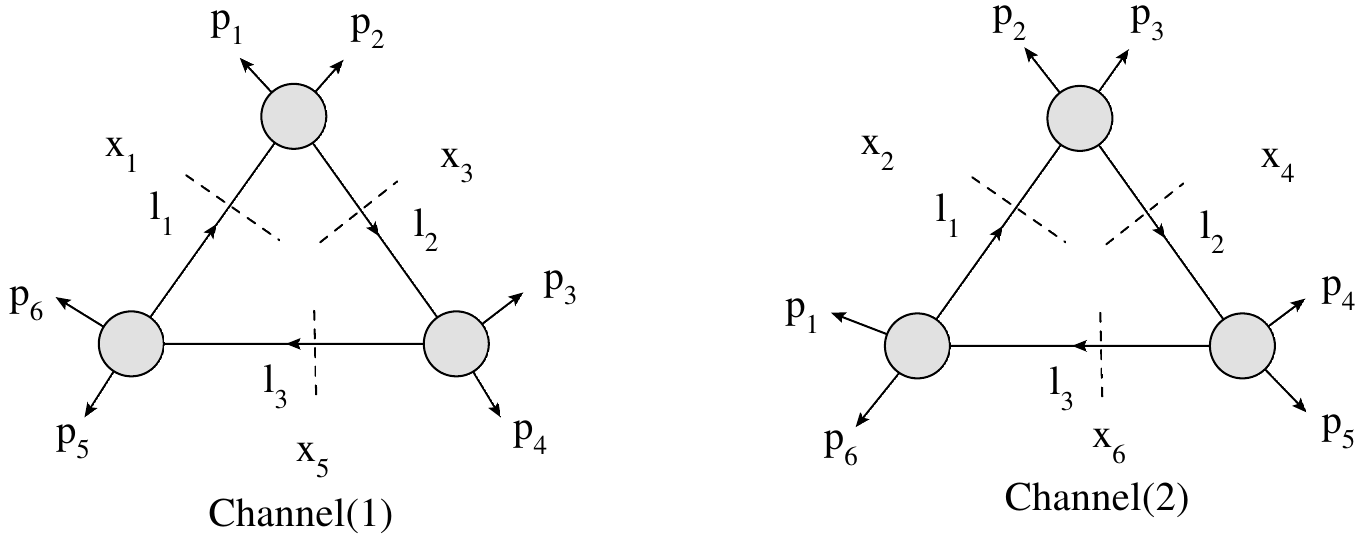}
\caption{The two maximal cuts at one-loop six-point. }
\label{ThreeCut}
\end{center}
\end{figure}

The other allowed scalar integrals are the massive triangles. Their coefficients along with that of the boxes can be fixed by the two triple-cuts $\mathcal{C}_{1,2}$ (and their conjugate $\mathcal{C}^*_{1,2}$), where the subscripts correspond to the the two distinct maximal cut, indicated as channel (1)  (2) in fig.~(\ref{ThreeCut}). Explicitly they are given by:
\eqa
\nonumber \mathcal{C}_{1}&=&\int \prod_{I=1}^3d\eta^I_{l_1}d\eta^I_{l_2}d\eta^I_{l_3}\mathcal{A}_4(\bar{1},2,\bar{l}_2,-l_1)\mathcal{A}_4(\bar{3},4,\bar{l}_3,-l_2)\mathcal{A}_4(\bar{5},6,\bar{l}_1,-l_3)\\
\nonumber \mathcal{C}_{2}&=&\int  \prod_{I=1}^3d\eta^I_{l_1}d\eta^I_{l_2}d\eta^I_{l_3}\mathcal{A}_4(-\bar{l}_1,2,\bar{3},l_2)\mathcal{A}_4(-\bar{l}_2,4,\bar{5},l_3)\mathcal{A}_4(-\bar{l}_3,6,\bar{1},l_1)
\eqae
We note that there is always an ambiguity in distinguishing $\mathcal{C}_1$ versus $\mathcal{C}_1^*$, since they arise from the two solutions of a quadratic equation.
However, two convention-independent combinations always exist. One is the average of the two cuts $\mathcal{C}_1+\mathcal{C}_1^*$ and the other is the average of the leading singularities, $LS_1+LS_1^* = (\mathcal{C}_1-\mathcal{C}_1^*)/[4\det (l_1,l_2,l_3)(\mathcal{C}_1)]$,
e.g., the numerators weighted by the Jacobian. Independence
of the second combination follows from sign flip of the Jacobian on the two solutions,
$\det(l_1,l_2,l_3)(\mathcal{C}_1) = -\det(l_1,l_2,l_3)(\mathcal{C}_1^*)$. 
The leading singularities have the following analytic form~\cite{Gang:2010gy}:
\eqa
 LS_1&=&\delta^3(P)\delta^6(Q)\frac{\prod^3_{I=1}(\alpha^{+I})}{2c^+_{25}c^+_{41}c^+_{63}}\;, \quad LS^*_1=LS_1(+\rightarrow-)\,.
\label{LSDef1}
\eqae
The functions $c^{\pm}_{ij}$ and $\alpha^{\pm I}$ are defined as 
\eqa
\nonumber c^{\pm}_{ij}:=\frac{\langle i|p_{135}|j\rangle\mp i \langle i+2,i-2\rangle \langle j-2,j+2\rangle}{p^2_{135}},\,\,\alpha^{\pm I}:=
 \frac{-(\epsilon_{\bar{i}\bar{j}\bar{k}}\langle \bar{i},\bar{j}\rangle\eta^I_{\bar{k}}\pm i\epsilon_{lmn}\langle l,m\rangle\eta^I_{n})}{p^2_{135}}\,,
\eqae
where in the definition of $\alpha^{\pm I}$, the (un-barred)barred indices indicate (odd)even labels. One can conveniently fix the convention of $\mathcal{C}_1$ and $\mathcal{C}^*_1$ as:
\eq
\mathcal{C}_1 :=2\l12\r\l34\r\l56\r LS_1, \quad \mathcal{C}_1^* = -\mathcal{C}_1(+\rightarrow -)\,.
\label{LSDef2}
\eqe
As one can check, the two combinations $LS_1+LS_1^*$ and $\mathcal{C}_1+\mathcal{C}_1^*$ both have the correct little group weights for an amplitude.

The leading singularities of ABJM have a dual presentation as the residues of an integral over orthogonal Grassmanian~\cite{Lee:2010du}. As discussed in ref.~\cite{Gang:2010gy} at $n=2k$-point there are $(k-2)(k-3)/2$ number of integration variables in the orthogonal Grassmanian. This implies that at six-point, there are no integrals to be done and one has a unique leading singularity from the Grassmanian (plus its complex conjugate due to the orthogonal condition). This implies that the second maximal cut $\mathcal{C}_2$ and $\mathcal{C}^*_2$ must be related to $\mathcal{C}_1$ and $\mathcal{C}^*_1$. Indeed one can check that 
\eq
\frac{\mathcal{C}_{1}+\mathcal{C}^*_{1}}{\langle12\rangle\langle34\rangle\langle56\rangle}=\frac{\mathcal{C}_{2}+\mathcal{C}^*_{2}}{\langle23\rangle\langle45\rangle\langle61\rangle}=-2i\mathcal{A}^\textrm{tree}_\textrm{6,shifted}\,,
\label{12Rel}
\eqe
where we have further identified the combination as the tree amplitude rotated by one,
$\mathcal{A}^\textrm{tree}_\textrm{6,shifted}(\bar{1}2\bar{3}4\bar{5}6):=\mathcal{A}^{\textrm{tree}}(\bar{2}3\bar{4}5\bar{6}1)$.
Note that all objects in this equation have the same little group weights (odd under reversal of the even $\lambda$'s)
so the identification makes sense.

A remarkable feature of 6-point kinematics is that the expressions for $\mathcal{C}_1$
are explicit in terms of angle brackets, that is they contain no square roots.
This reflects the fact that at six-points the cut solutions can be expressed explicitly in terms of angle brackets.
Let us see this explicitly.
At the same time, this will make apparent the following connection between the leading singularities and the BCFW form of the six-point tree amplitude,
\eq
\mathcal{A}^{\textrm{tree}}_{6}= LS_1+LS_1^*=\frac{\mathcal{C}_1-\mathcal{C}^*_1}{2\langle12\rangle\langle34\rangle\langle56\rangle} = LS_2+LS_2^*\,,
\label{TreeC}
\eqe 
in line with the original BCF logic \cite{Britto:2004ap} and as explained recently in \cite{Brandhuber:2012un}.
The main point is that the on-shell condition $l_1^2=l_2^2=0$ in channel (1) of fig.~(\ref{ThreeCut}) indicates that the loop momentum spinors can be parameterized as 
\eq
\lambda_{l_1}=\lambda_1\sin\theta+\lambda_2\cos\theta,\;\;\lambda_{l_2}=i(\lambda_1\cos\theta-\lambda_2\sin\theta)\,.
\label{sol}
\eqe
This is precisely the BCFW parameterization discussed in \cite{Gang:2010gy}.
On the double-cut there are three poles as a function of $\cos \theta$, whose residues are respectively $LS_1$, $LS_1^*$, and $-A_6^\textrm{tree}$. (The latter is located
at $\cos\theta=0$ and a computation of its residue is detailed in subsection~(\ref{TripleCutDis}), as part of our determination of the two-loop integrand.)
The desired relation then follows from the fact that the three residues must sum up to zero by Cauchy's theorem.
For completeness, we record here the explicit solution which corresponds to $\mathcal{C}_1$
\eq
\sin\theta = i c^+_{45}/\sqrt{(c^+_{36})^2-(c^+_{45})^2}  \quad \mbox{and}\quad  \cos\theta = c^+_{36}/\sqrt{(c^+_{36})^2-(c^+_{45})^2}.
\eqe
From eq.~(\ref{LSDef2}), one also sees that $\mathcal{C}_i$ has a non-uniform weight under conformal inversion:
\eq
I\left[\mathcal{C}_{1}\right]=\frac{\mathcal{C}_{1}}{\prod^6_{i=1}\sqrt{(x^2_i)}x^2_1x^2_3x^2_5},\;\;I\left[\mathcal{C}_{2}\right]=\frac{\mathcal{C}_{2}}{\prod^6_{i=1}\sqrt{(x^2_i)}x^2_2x^2_4x^2_6}\,.
\label{Invert}
\eqe

We are now ready to use the maximal cut to completely fix the integrand. 
Two types of integrals contribute to the cut in channel (1) in fig.~(\ref{ThreeCut}), the massive triangles as well as the ``two-mass-hard" box integrals.
As there are two solutions for the maximal cut in channel (1), giving different cut results $\mathcal{C}_1$ and $\mathcal{C}^*_1$, the massive triangle by itself cannot simultaneously reproduce  both. This implies the need for the box integrals. On the cut the box integrals give:
\eq
I_{box}(3,4,5,1)\bigg|_{\mathcal{C}_1}=\sqrt{2}\langle12\rangle\langle34\rangle\langle56\rangle,\quad
I_{box}(3,4,5,1)\bigg|_{\mathcal{C}^*_1}=-\sqrt{2}\langle12\rangle\langle34\rangle\langle56\rangle\,,
\label{BoxCut}
\eqe
where $|_{\mathcal{C}_{1}}$ indicates the maximal cut it is evaluated on.
A simple way to verify these formulas, up to a common sign, is to compare their square with the square of
$\epsilon(a,3,4,5,1)/(a\cdot 4)$ on the cut, using the identity
\eq
\epsilon(i_1,\ldots,i_5)\epsilon(j_1,\ldots,j_5) := \det \big[(i_i\cdot j_j)\big],
\label{note1}
\eqe
which in fact defines our normalization of the Levi-Cevita tensor.
The sign can be computed by a judicious use of eq.~(\ref{id1}).

Since the one-mass box must combine with the two-mass-hard box in the combination given in eq.~(\ref{1LoopBox}), this fixes the final integrand that reproduces the correct maximal cut to be (stripping a loop factor $4\pi N/k$):
\begin{equation}
\boxed{
\begin{split}
  \mathcal{A}_{6}^{\textrm{1-loop}}&=\frac{\mathcal{A}_{6}^{\textrm{tree}}}{\sqrt{2}} \bigg[I_{box}(3,4,5,1)+I_{box}(1,2,3,4)-I_{box}(4,5,6,1)-I_{box}(6,1,2,4)\bigg]\\[1ex]
    &{} +\frac{\mathcal{C}_{1}+\mathcal{C}^*_{1}}{2}I_{tri}(1,3,5)+\frac{\mathcal{C}_{2}+\mathcal{C}^*_{2}}{2}I_{tri}(2,4,6)\,.
\end{split}
}
\label{1LoopInt}
\end{equation}
Using eqs.~(\ref{TreeC}) and (\ref{BoxCut}), one can see that all maximal cut are correctly reproduced.

An important feature of the integrand in eq.~(\ref{1LoopInt}) is that it picks up a minus sign under a cyclic shift of the all scalar component amplitude by one-site. For the box integrals, this is a consequence of the linear combination dictated by the vanishing two-particle cut in eq.~(\ref{1LoopBox}). For the triangles, this is a consequence of their coefficients: If one considers the all scalar $\langle\bar{\phi}^4\phi_4\bar{\phi}^4\phi_4\bar{\phi}^4\phi_4\rangle$ component of the amplitude, from the explicit form of $\mathcal{C}_{1,2}$ in eqs.~(\ref{LSDef1})--(\ref{12Rel}), one sees that under a cyclic shift:
\eq
\left.\mathcal{C}_{1}(\bar{\phi}^4\phi_4\bar{\phi}^4\phi_4\bar{\phi}^4\phi_4)\right|_{i\rightarrow i+1}=-\mathcal{C}^{*}_{2}(\bar{\phi}^4\phi_4\bar{\phi}^4\phi_4\bar{\phi}^4\phi_4)\,.
\eqe
These additional signs are important for a non-vanishing one-loop amplitude as we now discuss. In ABJM, the tree and even-loop six-point amplitudes are parity even under $k\rightarrow-k$, while odd-loops are parity odd.
As parity is believed to be non-anomalous, this naively forbids non-trivial corrections from odd-loops unless these are odd under parity.
Due to the change from $k\rightarrow-k$, we are really exchanging the two gauge group U(N)$_{k}\times$U(N)$_{-k}$, and thus resulting in a cyclic shift in the identification of the barred and unbarred-multiplet. Thus if the one-loop amplitude picks up a minus sign under the cyclic shift, this will compensate for the parity odd nature, and gives an acceptable one-loop correction. This aspect of the one-loop amplitude has been discussed previously in ref.~\cite{Talk, Bianchi:2012cq,Bargheer:2012cp}.

\subsection{The one-loop amplitude}

The box integrals integrate to zero, thus the one-loop amplitude, at order $\mathcal{O}(\epsilon^0)$ is given solely by the massive triangles\footnote{The basic integral with massless internal lines, which follows easily from (\ref{intI0}) with $D=3$, is $\int_a \frac{1}{(a\cdot i)(a\cdot j)(a\cdot k)} = 1/[8 \sqrt{(i\cdot j)}\sqrt{(i\cdot k)}\sqrt{(j\cdot k)}]$.}:
\eqa
\nonumber\mathcal{A}_{6}^{\textrm{1-loop}}&=& \frac{N}{k}\left(\frac{\pi\left(\mathcal{C}_{1}+\mathcal{C}^*_{1}\right)}{4\sqrt{(1\cdot3)}\sqrt{(5\cdot3)}\sqrt{(1\cdot5)}}
+\frac{\pi\left(\mathcal{C}_{2}+\mathcal{C}^*_{2}\right)}{4\sqrt{(2\cdot4)}\sqrt{(4\cdot6)}\sqrt{(6\cdot2)}}\right)
\eqae
The fact that $(i\cdot i{+}2)=-\l ii{+}1\r^2$ motivates the following definition \cite{Bargheer:2012cp}:
\eq
 \sgnc\l ij\r:=\frac{\langle ij\rangle}{i\sqrt{- \l i j \r^2 -i\epsilon}}=\pm 1.
\eqe
Using eq.~(\ref{12Rel}) the one loop six-point ABJM amplitude can thus be rewritten as
\begin{equation}
\hspace{-0.2cm}
\boxed{
\begin{split}
  \mathcal{A}_{6}^{\textrm{1-loop}}&\!=\!\left(\frac{N}{k}\right)\frac{-\pi}{2}\mathcal{A}^\textrm{tree}_\textrm{6,shifted}
  \left(\sgnc\l12\r\sgnc\l34\r\sgnc\l56\r+\sgnc\l23\r\sgnc\l45\r\sgnc\l61\r\right).
  \end{split}\hspace{-0.3cm}}
\label{oneloopres1}
\end{equation}
Thus the one-loop amplitude is proportional to the tree-amplitude shifted by one-site multiplied by a step function.
This result has been obtained previously in \cite{Talk, Bianchi:2012cq,Bargheer:2012cp}.

In closing, we note that at six-point there are only two distinct Yangian invariant,
the sum and the difference of the leading singularity and it's conjugate. Interestingly, both combinations are local quantities, with the difference appearing as the tree-amplitude, while the sum appears as the one-loop amplitude. 
From eq.~(\ref{LSDef1}) this property is rather obscure, however due to the following non-trivial identity, equivalent to eq.~(\ref{id2}),
\eq
\l i|j+k|l\r^2+(p_i+p_j+p_k+p_l)^2\l jk\r^2=(p_i+p_j+p_k)^2(p_j+p_k+p_l)^2
\eqe
one finds for example 
\eq
c^+_{41}c^-_{41}=\frac{p^2_{345}}{p_{135}^2}\,.
\eqe
Thus the denominators of the leading singularities are in fact local propagators.\footnote{The locality of the leading singularities at six-point has been recently understood as a special property of the orthogonal Grassmaniann~\cite{Huang:2012vt}.} 
However, only the sum of the leading singularities has the correct little group weights to appear in an amplitude. The difference does not, unless it is multiplied by sign functions, which explains why it can appear only at loop level.


\subsection{Analytic properties of the one-loop amplitude} \label{sec:analysis}

The one-loop result (\ref{oneloopres1}) displays some remarkable properties which are worth spending some time on.
In particular, step functions are rarely seen in loop amplitudes, so we need to understand well why they are allowed to appear in three space-time dimensions.

First, we would like to give some topological interpretation to the region where the amplitude is nonzero.
In Minkowski space as null momenta can be parameterized as $p_i=E_i(1,\sin\theta_i,\cos\theta_i)$, the kinematic configuration of the scattering can be projected to a set of points on $S^1$.  The first thing to notice is that the invariant $\langle ij\rangle$ flips sign whenever the points $i,j$ on $S^1$ crosses each other,
as was also noted in \cite{Talk, Bianchi:2012cq,Bargheer:2012cp}. This is easy to see by writing the invariants in terms of coordinates on $S^1$:  
\eq
\langle ij\rangle=  2\sin\left(\frac{\theta_2-\theta_1}{2}\right) \sqrt{E_i+i\epsilon}\sqrt{E_j+i\epsilon}\,.
\eqe
Thus the function changes sign whenever point $j$ crosses point $i$ on $S^1$.
It is thus natural to divide the phase space into chambers depending on the ordering of the angles of the particles;
the one-loop amplitude is locally constant in each of these chambers.

Given that the product of sign functions changes sign whenever two angles cross, the angular dependence can be given a simple topological interpretation
in terms of a ``winding number'' counting the number of angle crossings compared to the color ordering.
This can be defined as follows: if multiples of $2\pi$ are added to angles such that they are strictly increasing, $0<\theta_{i+1}-\theta_i<2\pi$, $i=1\ldots7$,
then $w:=(\theta_7-\theta_1)/(2\pi)$.  Then one can show
\be
 \frac{\sgnc\l12\r\sgnc\l34\r\sgnc\l56\r}{\sgnc\l23\r\sgnc\l45\r\sgnc\l61\r} = (-1)^w (-1)^k.
\ee
The second factor $(-1)^k$ has a kinematical origin and originates from the factors $\sqrt{(i\cdot j)-i\epsilon}$ which can be real or imaginary depending on whether the given channel is space-like or time-like, respectively.  The number $k$ then simply equals the number of positive-energy timelike two-particle channels.
We see that the 1-loop amplitude is a highly intricate function of the kinematical configuration. 

As discussed in ref.~\cite{Bargheer:2012cp}, the fact that the one-loop amplitude is a step function can be readily understood from superconformal anomaly equations. Using free representation for the OSp(6$|$4) superconformal generators, it was shown that acting on the one-loop six-point amplitude with the linear generators, one must obtain an anomalous term that is proportional to $\delta(\langle ij\rangle)$, i.e. it has support on regions where two external legs become collinear. As the generators are linear, single derivatives in the on-shell variables, this implies that the amplitude must be proportional to step functions, or equivalently, sign functions.    

However, we are rather disturbed by the notion of an amplitude vanishing in an open set but nonzero elsewhere --- this would seem to clash with the amplitude being an analytic function of the external momenta.  In the rest of this section, we will propose that the step functions
behavior are not actually incompatible with analyticity of the amplitude, but are likely only an artifact of fixed-order perturbation theory.

\begin{figure}
\begin{center}
\eq\begin{array}{c@{\hspace{1.5cm}}c}
\includegraphics{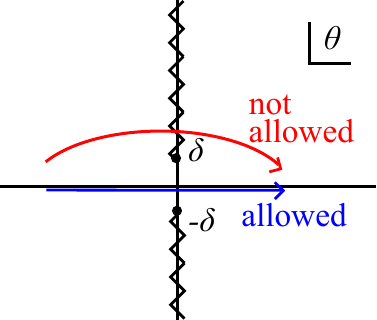}
&
\includegraphics{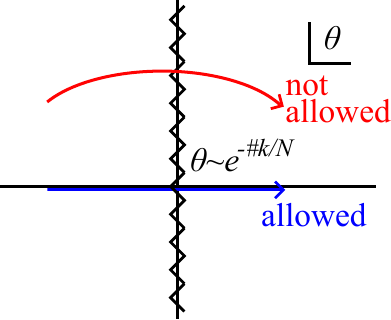}\\
\!\!\!\mbox{(a)}&\!\!\!\!\!\mbox{(b)}
\end{array}\nonumber\eqe
\caption{(a) The amplitude $F_\delta$ in the presence of a small mass.
It can be continued from $\theta<0$ to $\theta>0$ through a narrow window of size $\sim\delta$,
which shrinks to zero size in the massless limit.
(b) The advocated behavior in the massless setup, at a small but finite value of the coupling.  A branch cut covers the whole imaginary axis
but the discontinuity across it tends to zero at the origin.}
\label{fig:continuation}
\end{center}
\end{figure}

It is useful to first ask what would happen if we added small masses to the internal propagators still keeping the external lines massless.
This could arise naturally by giving a vacuum expectation value to some of the scalars of the theory as discussed in section~(\ref{sec:higgs}).
In that case, the sign function singularity would split into two threshold singularities
at $\theta=\pm \delta$ with $\delta=\sqrt{4m^2/E_1E_2}$.  Schematically,
\be
 \textrm{sgn}(\theta_2-\theta_1) \to F_\delta(\theta_2-\theta_1) 
\ee
where $F_\delta(\theta_2-\theta_1)$ is an analytic function with an analytic window of width $2\delta$ around the origin.\footnote{The precise form of $F_\delta$
can be worked out from the following exact expression for the internally massive loop integral,
writing $\Delta=\big(x_{ij}^2x_{ik}^2x_{jk}^2+m^2(2x_{ij}^2x_{ik}^2+2x_{ij}^2x_{jk}^2+2x_{ik}^2x_{ik}^2-x_{ij}^4-x_{ik}^4-x_{jk}^4)\big)^{1/2}$:
\eq
 \int_a \frac{1}{[(a\cdot i)+m^2][(a\cdot j)+m^2][(a\cdot k)+\mu^2]} = 
\frac{1}{8\pi i \Delta}
\log\frac{i\Delta + m(x_{ij}^2+x_{ik}^2+x_{jk}^2 +8m^2)}{-i\Delta + m(x_{ij}^2+x_{ik}^2+x_{jk}^2 +8m^2)}.
\eqe
In the collinear regime $x_{ij}^2\sim m^2$, this exhibits on the first sheet a pair of logarithmic branch points at the threshold $x_{ij}^2+4m^2=0$.
However, on the second sheet there is also a square-root branch point at $x_{ij}^2=m^2\frac{(x_{ik}^2-x_{jk}^2)^2}{x_{ik}^2x_{jk}^2}$.  The latter could be visible
with physical Minkowski space kinematics, depending on whether the $x_{ik}$ and $x_{jk}$ channels are time-like or not.
}
This amplitude is plotted in the complex $\theta$ plane in fig.~(\ref{fig:continuation}). We see that as long as $m\neq 0$
there exists a small window of width $\delta$ around the origin along which the amplitude can be rightfully continued.

We also see clearly why such behavior is possible in three space-time dimensions but not in higher dimensions.
In three dimensions the physical (real) phase space for a set of massless particles splits into chambers which are separated by singular, collinear configurations.
To analytically continue from one chamber to the next one must avoid the singularity, since the amplitude is not required to be analytic around that point.
But attempts to avoid the singularity by passing through the complex plane may fail: the singularity can be surrounded by cuts.

At the massless point but at the nonperturbative level,
we expect an analogous situation but with a nonperturbatively small window of width $\delta\sim e^{-\#\frac{k}{N}}$.
Indeed, in a theory where soft and collinear quanta are copiously produced, as is ABJM,
we find it unlikely for a sharp feature such as a sign function to remain unwashed.
Rather, the backreaction of the radiation on the ongoing hard quanta should smear the small angle behaviour.
In perturbation theory this would become visible through large logarithms $N/k \log 1/\theta$, which would have to be resumed at small angles.
Indeed such logarithms will come out of our two-loop computation.
Thus a more faithful model for the small angle behavior at small but finite coupling should be a function of the sort
\be
 \textrm{sgn}(\theta_2-\theta_1) \to \frac{\theta_2-\theta_1}{( (\theta_2-\theta_1)^2)^{\frac12-\#N/k}}
\ee
which can be happily continued from the left region to the right region. It would be very interesting to investigate the small-angle behavior quantitatively and confirm
that the discontinuity across the cut goes to zero as $\theta_2-\theta_1\to 0$.

\section{The two-loop six-point integrand}
\label{sec:integrand}
We shall now proceed to determine the two-loop six-point integrand from a variety of on-shell constraints.
In ABJM theory we get a large number of constraints just from
the fact that there are no 3- and 5-point on-shell amplitudes.
This gives a large number of cuts on which the integrand must vanish.
In addition, there are some very simple non-vanishing triple-cuts associated with soft gluon exchanges which can be used to fix the remaining
freedom.

\begin{figure}
\begin{center}
\includegraphics{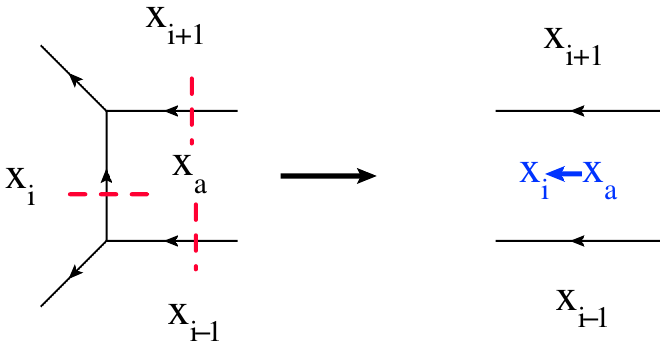}
\caption{The triple cut of consecutive massless corners corresponds to soft exchange between the two external lines. In dual space, this correspond to the loop region $x_a$ approaching $x_i$. }
\label{3ptcut}
\end{center}
\end{figure}

Our first goal in this section is thus to determine the two-loop hexagon integrand using just
the following constraints:
\begin{itemize}
\item[0.] The integrand is dual conformal invariant.
\item[1.] Cuts isolating a five-point amplitude must vanish.
\item[2.] Cuts isolating a three-point vertex must vanish.
\item[3.] Triple-cuts of consecutive massless corners, as shown in fig.~(\ref{3ptcut}), correspond to soft gluon exchange and must reduce to the 1-loop integrand.
\item[4.] Absence of non-factorizable collinear divergences.
\end{itemize}

\begin{figure}
\begin{center}
\includegraphics{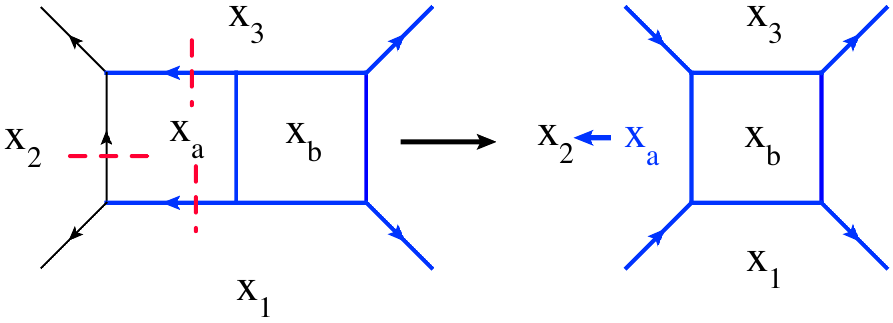}
\caption{The triple cut of the double box integral. As $x_a$ approaches $x_2$ on the cut condition, one should obtain the one-loop integrand given in eq.~(\ref{1loopBox}).}
\label{3ptcut2}
\end{center}
\end{figure}

As an example, we now show that by simply using steps $2$ and $3$, one completely fixes the four-point two-loop integrand to be that constructed in ref.~\cite{Chen:2011vv}. This also illustrate the importance of obtaining the one-loop amplitude beyond $\mathcal{O}(\epsilon)$. The one-loop four-point integrand is given by 
\eq
\frac{\mathcal{A}_4^{\textrm{tree}}}{\sqrt{2}}\frac{\epsilon(a1234)}{(a\cdot1)(a\cdot2)(a\cdot3)(a\cdot4)},
\label{1loopBox}
\eqe
where the unpleasant-looking factor of $\sqrt{2}$ is due to our normalization of the five-dimensional $\epsilon$-symbol as discussed around eq.~(\ref{note1}).
Now consider the triple cut of a double-box integral in fig.~(\ref{3ptcut2}). On the cut, $x_a$ approaches $x_2$ and in this limit one should recover eq.~(\ref{1loopBox}). With a little thought one sees that the following double-box numerator does the job:
\eq
\frac{\mathcal{A}_4^{\textrm{tree}}}{2}\frac{\epsilon(a123*)(b341*)}{(a\cdot1)(a\cdot2)(a\cdot3)(a\cdot b)(b\cdot 3)(b\cdot 4)(b\cdot 1)}\,
\eqe
where $\epsilon(a,i,j,k,*)\epsilon(b,l,m,n,*):=\epsilon(a,i,j,k,\,^\mu)\epsilon(b,l,m,n,\,_\mu)$. The detailed behavior of such numerator under the cut condition will be discussed in subsection~(\ref{TripleCutDis}). This however, is not complete as one sees that there is a non-trivial contribution to the cut $(a\cdot3)=(a\cdot b)=(b\cdot 3)=0$. This separates out a three-point tree amplitude and hence must vanish.
On this cut, using (\ref{note1}) and setting $y_a=y_b$ to restrict to an easy subcase, the double box gives a nontrivial contribution 
\eq
-\frac{\mathcal{A}_4^{\textrm{tree}}}{2}\frac{(1\cdot3)^2}{(a\cdot1)(b\cdot 1)}\,.
\eqe
One can easily see that this contribution can be cancelled by a double triangle integral. Thus combining requirements $(2)$ and $(3)$ uniquely fixes the two-loop four-point integrand to be:
\eq
\nonumber\mathcal{A}_{4}^{\textrm{2-loop}}=\frac{\mathcal{A}_4^{\textrm{tree}}}{2}\int_{a,b}\left[\frac{\epsilon(a123*)(b341*)+(a\cdot 2)(b\cdot 4)(1\cdot 3)^2}{(a\cdot1)(a\cdot2)(a\cdot3)(a\cdot b)(b\cdot 3)(b\cdot 4)(b\cdot 1)} \;\;+\;\;(s\leftrightarrow t)\right].
\eqe
One can see that the above also satisfy requirement (1) and is equivalent to that of~\cite{Chen:2011vv}. 

\subsection{Integrand basis}
We begin by constructing the most general algebraic basis of dual-conformal integrals at two loops.
In three dimensions, the most general two-loop integral is a double-box
\be
 I_{2box}^{ijk;lmn}[(v_1\cdot a)(v_2\cdot b)] :=
  \int_{a,b} \frac{(v_1\cdot a)(v_2\cdot b)}{(a\cdot i)(a\cdot j)(a\cdot k)(a\cdot b)(b\cdot l)(b\cdot m)(b\cdot n)} \nonumber
\ee
where $v_1$ and $v_2$ are some 5-vectors.
Note that the presence of the numerator is required by dual conformal invariance,
as the integrand must have scaling weight -3 with respect to both $a$ and $b$.%
\footnote{The absence of pentagon-boxes or more complicated topologies can be easily proved as follows.
 A pentagon would need a numerator quadratic in $a$.
 Let's assume the five external propagators involving $a$ are $a_1,\ldots a_4$ and $a_b$. Then  we can expand the numerator
 in terms of products $(a\cdot v_1)(a\cdot v_2)$
 where the $(a\cdot v_i)$ are chosen lie in the following basis
 \be
  (a\cdot 1),(a\cdot2),(a\cdot3),(a\cdot 4) \;\; \mbox{and}\;\; \epsilon(a1234).
 \ee
 All numerators in this basis trivially cancel some propagator, except for $(\epsilon(a1234))^2$, which would appear to be irreducible.
 However, this can be reduced using the Gram identity (\ref{note1}) together with $a^2=0$.}
Numerators $v_1$ proportional to $y_i, y_j$ or $y_k$ are reducible, which would leave a-priori 2 distinct numerators on each side.
However, at 6 points constraint 2 above is very powerful as it requires the numerator to have zeros on any double cut isolating a massless external leg.
For dual conformal invariant integrals, this restricts the numerators to be of the $\epsilon$-type
\eq
\nonumber I_{2box}^{ijk;lmn}[\epsilon(a,i,j,k,*)\epsilon(b,l,m,n,*)] \quad\mbox{or}\quad I_{2box}^{ijk;kli}[\epsilon(a,i,j,k,b)]\,,
\eqe
where the second possibility is allowed only when $(k\cdot l)$ and $(j\cdot l)$ are both nonvanishing.
Note that this latter parity-odd double-box integral has excessive weight on $(i,k,l)$. At six-point, this can be naturally absorbed by the extra weights of $\mathcal{C}_1+\mathcal{C}_1^*$ shown in eq.~(\ref{Invert}). 

\begin{figure}
\begin{center}
\includegraphics[scale=0.9]{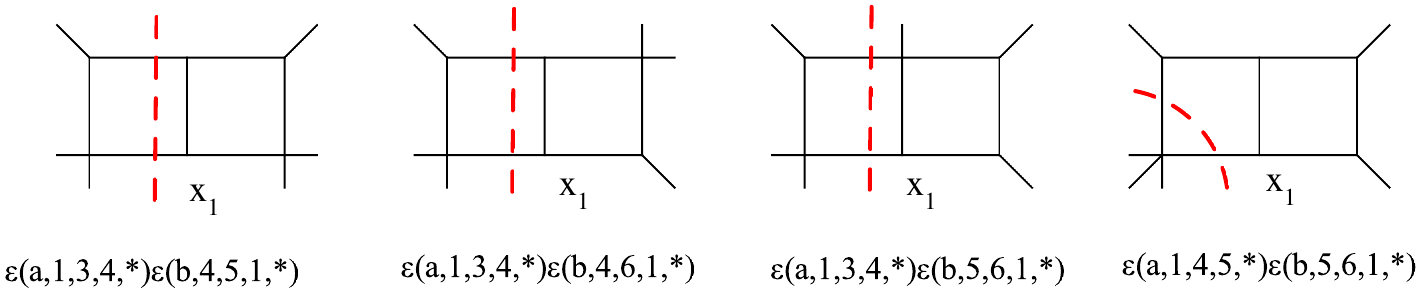}
\caption{Possible box integrals that have non-trivial two-particle cut which would correspond to factorization channel that factorize the amplitude into a product of 5-pt amplitudes. The contributions to the cut from each box integral are distinct, leading to the conclusion that they will not appear.}
\label{5pt}
\end{center}
\end{figure}

Six-point double-box integrals with a three-legged massive corner, and some with two legged massive corners, will have two-particle cuts that factories into a product of five-point amplitudes as shown in fig.~(\ref{5pt}). Since five-point amplitudes vanish to all order in $\epsilon$, the contributions of these double-box integrals must cancel out on such cuts or they are not allowed in the integral basis. It is straight forward to see that the contributions are distinct and cannot cancel.
Thus by imposing conditions 1 and 2 on one-loop subdiagrams the allowed parity even double box integrals are restricted to 
\eqa
\nonumber I_{even}^{\rm 2mh}(i)&:=&
  \int_{a,b} \frac{\epsilon(a,i,i+1,i+2,*)\epsilon(b,i+2,i-2,i,*)}{(a\cdot i)(a\cdot i+1)(a\cdot i+2)(a\cdot b)(b\cdot i+2)(b\cdot i-2)(b\cdot i)}\\
\nonumber I^{\rm crab}(i)&:=&
  \int_{a,b} \frac{\epsilon(a,i,i+1,i+2,*)\epsilon(b,i-2,i-1,i,*)}{(a\cdot i)(a\cdot i+1)(a\cdot i+2)(a\cdot b)(b\cdot i-2)(b\cdot i-1)(b\cdot i)}\\  
\nonumber I^{\rm critter}(i)&:=&
  \int_{a,b} \frac{\epsilon(a,i,i+1,i+2,*)\epsilon(b,i+3,i+4,i+5,*)}{(a\cdot i)(a\cdot i+1)(a\cdot i+2)(a\cdot b)(b\cdot i+3)(b\cdot i+4)(b\cdot i+5)}\\
  I_{odd}^{\rm 2mh}(i)&:=&
  \int_{a,b} \frac{\epsilon(a,i,i+1,i+2,b)}{(a\cdot i)(a\cdot i+1)(a\cdot i+2)(a\cdot b)(b\cdot i+2)(b\cdot i-2)(b\cdot i)} \label{defI6}
\eqae
where the subscript {\it even} and {\it odd} denotes the two-mass hard integrals with parity-even and -odd numerators.  

The same conditions also leave box-triangle integrals
\be
 I_{box;tri}^{ijk;lm}[\epsilon(a,i,j,k,l)] :=  \int_{a,b} \frac{\epsilon(a,i,j,k,l)}{(a\cdot i)(a\cdot j)(a\cdot k)(a\cdot b)(b\cdot l)(b\cdot m)}.
\ee
Other choices for the numerator here, such as the other natural choice $\epsilon(a,i,j,k,m)$, would be related by a Schouten identity plus double triangle integrals.
The box-triangles again have excessive weight which would imply that they should come with factors of $\mathcal{C}_1+\mathcal{C}_1^*$. We will see that they indeed arise in this way.

Finally, the conditions applied so far leave only three double-triangle integrals, namely
\eqa
\nonumber I_{2tri}^{i,i+2;i+2,i}&:=&\int_{a,b} \frac{(i\cdot i+2)^2}{(a\cdot i)(a\cdot i+2)(a\cdot b)(b\cdot i)(b\cdot i+2)}\\
\nonumber I_{2tri}^{i,i+2;i-2,i}&:=&\int_{a,b} \frac{(i\cdot i+2)(i\cdot i-2)}{(a\cdot i)(a\cdot i+2)(a\cdot b)(b\cdot i-2)(b\cdot i)}\\
  I_{2tri}^{i,i+2;i-3,i-1}&:=&\int_{a,b} \frac{(i\cdot i+2)(i-1\cdot i-3)}
{(a\cdot i)(a\cdot i+2)(a\cdot b)(b\cdot i-3)(b\cdot i-1)}. \label{doubletris}
\eqae
We now finish to implement constraint 2, the vanishing of all three-point sub amplitudes.

\subsection{Constraints from vanishing three point sub amplitudes}
We consider the cut $(a\cdot 3)=(a\cdot b)=(b\cdot 3)=0$ which separates out a three-point amplitude and thus must vanish.
Two types of double boxes contribute to such cut, $I^{2mh}$ and $I^{crab}$, and they contribute:
\eqa
\nonumber(1)&&\\
\nonumber&&\includegraphics[scale=0.8]{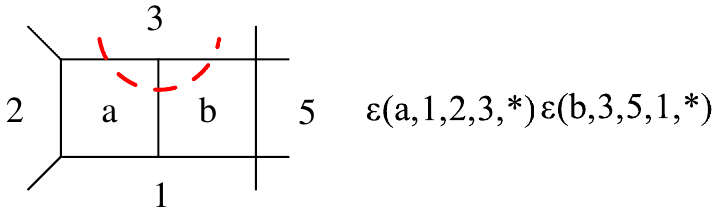}\\
\nonumber &\rightarrow& \frac{(1\cdot3)(b\cdot 2)[(a\cdot 1)(3\cdot 5)-(a\cdot 5)(3\cdot 1)]}{(a\cdot2)(a\cdot1)(b\cdot1)(b\cdot5)}\\
\nonumber(2)&&\\
\nonumber&&\includegraphics[scale=0.8]{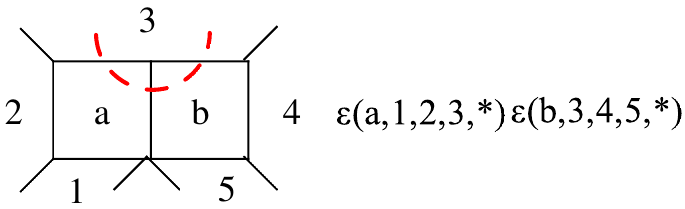}\\
\nonumber &\rightarrow&-\frac{(b\cdot2)(1\cdot3)(a\cdot4)(3\cdot5)}{(a\cdot1)(a\cdot2)(b\cdot5)(b\cdot4)}\\
\nonumber(3)&&\\
\nonumber&&\includegraphics[scale=0.8]{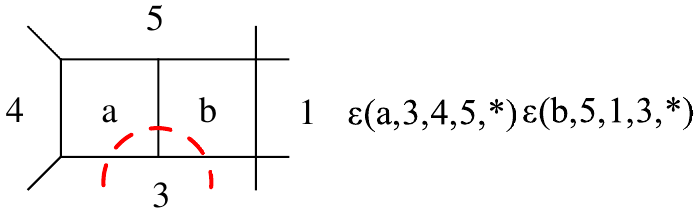}\\
\nonumber &\rightarrow& \frac{(b\cdot4)(3\cdot5)[(a\cdot5)(3\cdot1)-(a\cdot1)(3\cdot5)]}{(a\cdot4)(a\cdot5)(b\cdot5)(b\cdot1)}\\
\nonumber(4)&&\\
\nonumber&&\includegraphics[scale=0.8]{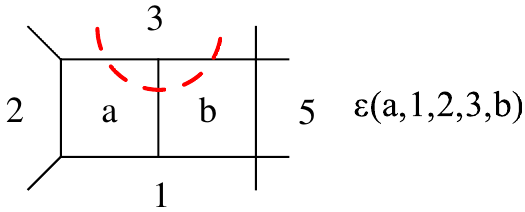}\rightarrow0\,,
\eqae 
where we've indicated the non-vanishing remainder on the cut.
This was obtained by using eq.~(\ref{note1}) to reduce the $\epsilon$-symbols to dot products and dropping terms which vanish on the cut.
This can be simplified further when we take into account that the general solution to the cut is parametrized by $y_{a,b}=y_3+\tau_{a,b} v$, where $v$ is any null five-vector such that $v{\cdot}3=0$. Physically, on the cut the two loop momenta are collinear with each other.
Then one finds that the following combination of double-box and double-triangle integrals vanish on the cut and are thus allowed
\eqa
\nonumber&&I_{even}^\textrm{2mh}(1)+ I_{2tri}^{1,3;3,1}- I_{2tri}^{1,3;3,5}-I_{2tri}^{1,3;5,1}\\
\nonumber&&I^{\rm crab}(1)+ I_{2tri}^{1,3;3,5}\\
\nonumber&& I^\textrm{critter}, \quad I_{2tri}^{i,i+2;i-1,i-3},\quad  I_{box;tri}^{i,i+1,i+2;i-1,i-3}\quad \mbox{and}\quad I_{odd}^\textrm{2mh}\,.
\eqae 
In addition the integral $I_{box;tri}^{i,i+1,i+2;i+2,i-2}$ is immediately ruled out.
In the following, we will use constraint 3 to fix the relevant coefficient of the double box integrals. 

\subsection{Constraints from one-loop leading singularity\label{TripleCutDis}}
The particular cut we will be interested in is the maximal cut of one of the sub loops with adjacent massless legs. This cut corresponds to a kinematic configuration where there is a soft-exchange between the two external legs, as can be deduced from eq.~(\ref{sol}) using the fact that the $(a\cdot 2)$ only gives a pole at $\cos\theta\to 0$.
In terms of dual regions, this correspond to when the loop region $y_a\rightarrow y_i$ as illustrated in fig.~(\ref{3ptcut}).

More specifically, we compute the leading singularity $(a\cdot1)=(a\cdot2)=(a\cdot3)=0$ of
$$\nonumber \frac{\epsilon(a,1,2,3,*)}{(a\cdot 1)(a\cdot 2)(a\cdot 3)(a\cdot i)}.$$
Normally there are two solutions to such a cut constraint, but let us verify explicitly that here there is only one solution $y_a=y_2$ as claimed.
To do so we expand $a$ over a natural basis, such as $a=a_1y_1+y_2+a_3y_3+a_4y_4+a_\epsilon y_\epsilon$ where $y_\epsilon:=\epsilon(1,2,3,4,*)$.
Imposing the two cuts $(a\cdot1)=(a\cdot 3)=0$ gives that $a_1=0$ and $a_3/a_4=-(1\cdot 4)/(1\cdot 3)$, and thus $a_3\propto a_\epsilon^2$ due to the $y^2=0$ constraint. Then $(a\cdot 2)\sim a_\epsilon^2$ so the only solution is $a_\epsilon=0$.

Let us work out the details. Normalizing the leading singularities in a convenient way
\eq
 \frac{F(a)}{(a\cdot i)(a\cdot j)(a\cdot k)}\bigg|_{\substack{\textrm{residue}\\i,j,k}} := 4\int_a \delta((a\cdot i)\delta((a\cdot j))\delta((a\cdot k))F(a),
\eqe
we have here
\eq
 4\int_a=4\int \frac{d^5a~\delta(a^2)}{\textrm{vol(GL(1))}} = N\int da_1da_3da_4da_\epsilon~\delta(a^2)  \nonumber
\eqe
where $N=\sqrt{2}\epsilon(y_1,y_2,y_3,y_4,y_\epsilon)=\sqrt{2}y_\epsilon^2$.
After taking the first two cuts and evaluating the Jacobian from the $\delta$-functions we get
\eq
 \int_a \frac{\delta((a\cdot 1))\delta((a\cdot 3))}{(a\cdot 2)} = \frac{N}{y_\epsilon^2(1\cdot3)^2(2\cdot 4)}\int \frac{da_\epsilon}{a_\epsilon^2} \label{doublepole}
\eqe
where $a(\epsilon)=y_2+a_\epsilon y_\epsilon+ a_\epsilon^2\frac{y_\epsilon^2}{2(2\cdot4)} \big[\frac{(1\cdot4)}{(1\cdot 3)}y_3-y_4\big]$
(this could be mapped to the BCFW parametrization (\ref{sol}), since this solves the same cut constraints).
Multiplying by $\epsilon(a,1,2,3,*)/(a\cdot i)$ and taking the residue at $a_\epsilon=0$, we immediately get the box leading singularity
\eq
\frac{\epsilon(a,1,2,3,*)}{(a\cdot 1)(a\cdot 2)(a\cdot 3)(a\cdot i)} \bigg|_{\substack{\textrm{residue}\\1,2,3}}
= \sqrt{2}\left[\frac{(2\cdot *)}{(2\cdot i)}\right]\,.
\eqe 
Note that although the numerator vanishes on the cut solution $y_a=y_2$, reflecting the
absence of three-point vertices in this theory, a nonvanishing residue remains due to the double pole in (\ref{doublepole}).  The residue reflects the
the exchange of a zero-momentum Chern-Simons field.

This physical origin implies that these leading singularities are ``universal'', and must reduce to
the lower-loop integrand with the loop variable $y_a$ omitted.  Thus, with a normalization easily fixed from the 1-loop integrand,
\eq
 \mathcal{A}^{\ell-\textrm{loop}}_n\bigg|_{\substack{\textrm{residue}\\1,2,3}} = \mathcal{A}^{(\ell-1)-\textrm{loop}}_n.
\eqe
Similarly, but with an opposite sign due to $k\to -k$,
\eq
 \mathcal{A}^{\ell-\textrm{loop}}_n\bigg|_{\substack{\textrm{residue}\\2,3,4}} = -\mathcal{A}^{(\ell-1)-\textrm{loop}}_n.
\eqe
These relations can easily be verified to hold for the one-loop integrand (\ref{1LoopInt}), where the right-hand side reduces to the tree amplitude.
However, these relations must hold at any loop order.
In a sense they are analogous to the so-called rung rule \cite{Bern:1997nh}.

\begin{figure}
\begin{center}
\includegraphics[scale=0.95]{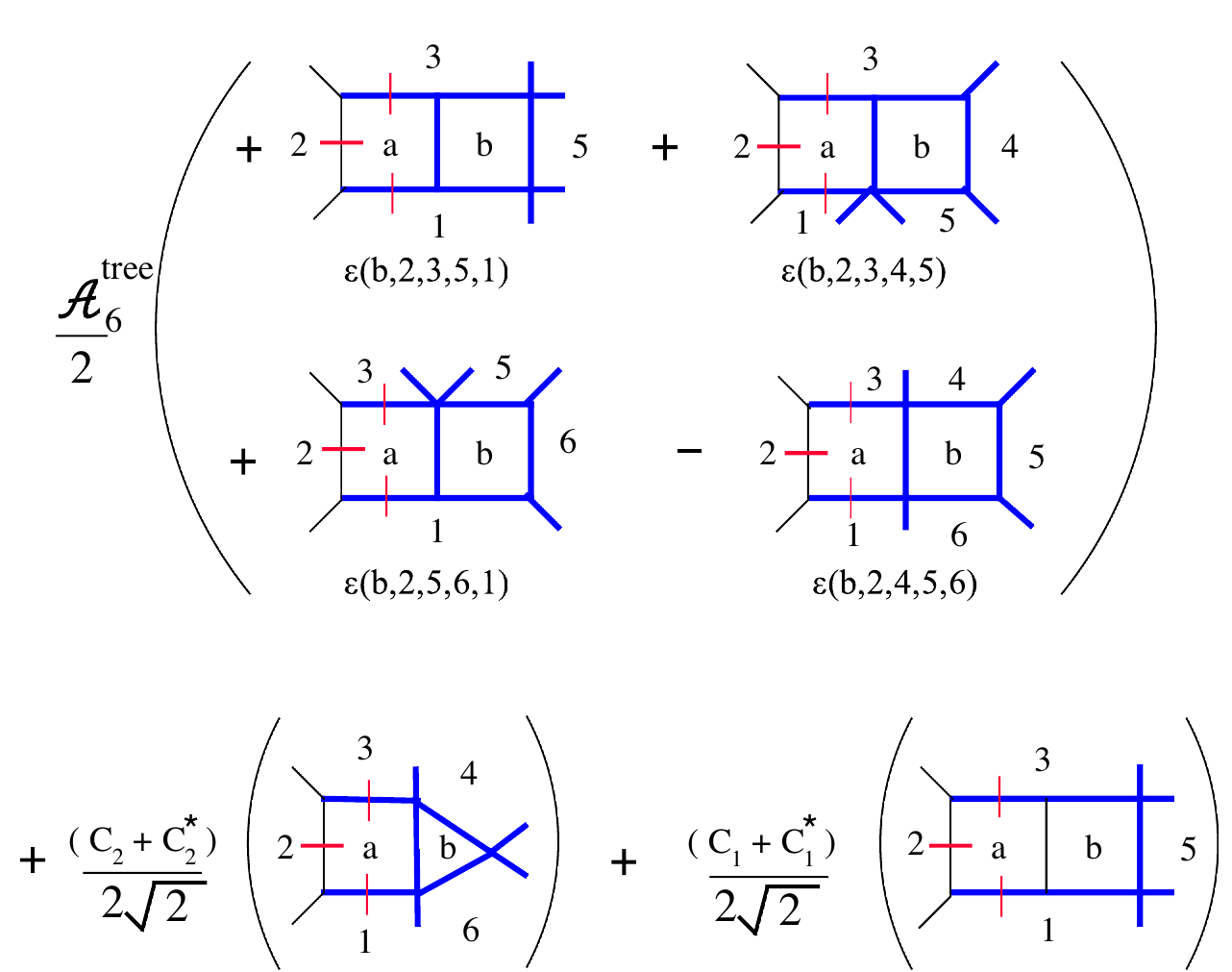}
\caption{The terms that contribute to the leading singularity $(a\cdot 1)=(a\cdot 2)=(a\cdot 3)=0$, which has to reduce to the one-loop integrand.
The blue lines indicate the one-loop propagators that remain and uncancelled after the cut. The term in the bottom of each diagram is the numerator factor. One can see that the combination is precisely the one-loop answer.}
\label{LeadingSin}
\end{center}
\end{figure}

At two loops, this requires to see the one-loop integrand emerge on the cut, i.e. eq.~(\ref{1LoopInt}).
Indeed one finds that the various pieces of the one-loop integrand do appear from the double-box and the box-triangle integrands. More specifically, for the cut $(a\cdot1)=(a\cdot2)=(a\cdot3)=0$, omitting the common $\sqrt{2}$ factor, we find the following contributions: 
\eqa
\nonumber I_{even}^{\textrm{2mh}}(1)&\rightarrow& I_{box}(3,5,1,2)\\
\nonumber I^{\textrm{crab}}(1)&\rightarrow& I_{box}(5,6,1,2),\;\;I^{\textrm{crab}}(3)\rightarrow I_{box}(3,4,5,2)\\
\nonumber I^{\textrm{critter}}(1)&\rightarrow& I_{box}(4,5,6,2),\\
I_{box;tri}^{1,2,3;4,6}[\epsilon(a,1,2,3,i)]&\rightarrow& (2\cdot i)I_{tri}(2,4,6),\;\;I_{odd}^{\textrm{2mh}}(1)\rightarrow I_{tri}(1,3,5)\,.
\eqae
As one can see, all one-loop integrals appearing in eq.~(\ref{1LoopInt}) are present. Thus the constraint of reproducing the one-loop integrand on the one-loop maximal cut, combined with previous results derived from constraint 2, fixes the two-loop integrand to be the following combination:
\eqa
\nonumber&&\frac{\mathcal{A}_{6}^{\textrm{tree}}}{2}\bigg[I_{even}^{\textrm{2mh}}(1)+I^{\textrm{crab}}(1)-I^{\textrm{critter}}(1)+I_{2tri}^{1,3;3,1}-I_{2tri}^{1,3;5,1}+{\rm cyclic}\bigg]\\
\nonumber&&+\frac{\mathcal{C}_{1}+\mathcal{C}^*_{1}}{2\sqrt{2}}\bigg[ I_{odd}^{\textrm{2mh}}(1)-\frac{I_{box;tri}^{4,5,6;1,3}[\epsilon(a,4,5,6,1)]}{(1\cdot 5)}+{\rm cyclic\times2}\bigg]\\
\nonumber&&+\frac{\mathcal{C}_{2}+\mathcal{C}^*_{2}}{2\sqrt{2}}\bigg[ -I_{odd}^{\textrm{2mh}}(2)+I_{box;tri}^{1,2,3;4,6}[\epsilon(a,1,2,3,6)]/(2\cdot 6)+{\rm cyclic\times2}\bigg]\\
\nonumber&&+\sum_{i=1}^6\alpha_i I_{2tri}^{i,i+2;i-3,i-1}
\label{Almost}
\eqae
where ${\rm cyclic\times2}$ implies cyclic by two sites and $\mathcal{C}_{1,2}, \mathcal{C}^*_{1,2}$ are defined as before. The presence of the one-loop integrand on the cut $(a\cdot1)=(a\cdot2)=(a\cdot3)=0$ are shown in fig.~(\ref{LeadingSin}).
We see that at this point, the only remaining freedom is the triangle integrals $I_{2tri}^{i,i+2;i-3,i-1}$.
However, as we will now see these integrals are ``badly'' collinear divergent and so they are constrained by other physical considerations.

\subsection{Collinear divergences and the ABJM two-loop integrand}
As was demonstrated in \cite{Collinear} (in the context of planar $\mathcal{N}=4$),
the exponentiation of divergences leads to constraints which can be formulated in a very simple way at the level of the integrand,
e.g., before even performing any integral.
We will now formulate similar constraints in ABJM theory, but these will have a somewhat different flavor due to the absence of one-loop divergences.

In ABJM theory, the twist-two anomalous dimensions which control the collinear and soft-collinear divergences
begin at order $(k/N)^2$, e.g. two-loops. Thus the divergences at two-loop are the leading ones and
must be proportional to the tree amplitude in a specific way.
Some qualitative constraints can be deduced in a simple way as follows:
We note that soft divergences can be computed by replacing the external states by Wilson lines.
Just this fact imposes two simple constraints.
First, the coefficient of proportionality of the $1/\epsilon^2$ divergence must be a pure number, e.g. independent of the kinematics (ultimately, 6 times the so-called cusp anomalous dimension).  Second, kinematic dependence of the subleading $1/\epsilon$ divergence, which can arise from soft wide-angle radiation but not collinear radiation (and hence is controlled by the Wilson lines) can only be of the simple ``dipole'' invariants of the form $[1/\epsilon]\log \frac{x_{i,i{+}1}^2}{\mu^2_\textrm{IR}}$.
We will call divergences of these forms ``factorizable''.
These are rather general constraints that any physically acceptable amplitude must possess
and we will see that they impose nontrivial constraints on the integrand.

We will consider the collinear divergence from the region collinear to momentum $p_3$. To have a divergence we need both loop momenta to be collinear, thanks to the
special $\epsilon$-numerators, so we consider the limit
\eq
 y_a\rightarrow y_3 + \tau_a y_4,\;\; y_b\rightarrow y_3 + \tau_b y_4.
 \label{Colinear}
\eqe

A first requirement is that the integrals proportional to the parity-odd structure, e.g. $\mathcal{C}_i+\mathcal{C}_i^*$, be finite.
For the box-triangles, we find the following combination is free of divergences:
\eqa
\nonumber&&(\mathcal{C}_{1}+\mathcal{C}^*_{1})\bigg(\frac{I_{box;tri}^{4,5,6;1,3}[\epsilon(a,4,5,6,1)]}{(1\cdot 5)}\bigg)-(\mathcal{C}_{2}+\mathcal{C}^*_{2})\bigg(\frac{I_{box;tri}^{1,2,3;4,6}[\epsilon(a,1,2,3,6)]}{(2\cdot 6)}\bigg).\nonumber
\eqae
To see that this combination is indeed finite in the collinear region, note that in the limit eq.~(\ref{Colinear}), factoring out the divergent factors $1/(a\cdot3) (a\cdot b)( b\cdot 4)$ one has 
\eqa
\nonumber&&(\mathcal{C}_{1}+\mathcal{C}^*_{1})\frac{\epsilon(3,4,5,6,1)}{(5\cdot 1)(a\cdot1)(3\cdot5)(b\cdot6)}-(\mathcal{C}_{2}+\mathcal{C}^*_{2})\frac{\epsilon(4,1,2,3,6)}{(2\cdot 6)(a\cdot1)(4\cdot2)(b\cdot6)}\,.
\eqae
where we've symmetrized in $(a\leftrightarrow b)$. The above combination vanishes thanks to the following identity:
\eq
\frac{\mathcal{C}_{1}+\mathcal{C}^*_{1}}{\mathcal{C}_{2}+\mathcal{C}^*_{2}} =-\frac{\epsilon(6,1,2,3,4)(3\cdot5)(5\cdot1)}{\epsilon(3,4,5,6,1)(6\cdot2)(2\cdot4)}\,. 
\label{Id}
\eqe 
This identity is proven in appendix~(\ref{IdApp}).  Thus we conclude that the parity-odd part of the integrand in eq.~(\ref{Almost}) is already complete, provided
that the box-triangle numerators are chosen as there.

We now turn to the parity-even sector.  We need to study the divergences of the yet-unconstrained integral $I_{2tri}^{1,3;4,6}$ in more detail.
Integrating out the remaining variables around the limit (\ref{Colinear}) one easily obtains the divergent contribution from the collinear region
\eq
 \int_{a,b} \frac{1}{(a\cdot 3)(a\cdot b)(b\cdot 4)(a\cdot 1)(b\cdot6)}  \propto \log \mu^2\int_{0<\tau_a<\tau_b<\infty} \frac{d\tau_a d\tau_b}{\sqrt{\tau_a (\tau_b-\tau_a)}(a\cdot 1)(b\cdot6)} \nonumber
\eqe
where $a$ and $b$ are as in (\ref{Colinear}).
Such a divergence violates factorizability in \emph{two} ways: it depends on $y_1$ through $(a\cdot 1)$ \emph{and} on $y_6$ through $(b\cdot 6)$.
This leads, for instance, to dependence on the cross-ratio $u_1$.
A quick look at (\ref{Almost}) reveals that the only other integral with potentially similar dependence on $\tau_{a,b}$ is $I^\textrm{critter}$.
However the divergence cancels exactly, pre-integration, in the combination
\eq
I^{\textrm{critter}}(1)+I_{2tri}^{1,3;4,6}\,. \nonumber
\eqe
Thus we finally arrive at the complete integrand for two-loops six-point amplitude in ABJM theory:
\begin{equation}
\boxed{
\begin{split}
    \mathcal{A}_{6}^{\textrm{2-loop}} &=\left(\frac{4\pi N}{ k}\right)^2\bigg\{\frac{\mathcal{A}_{6}^{\textrm{tree}}}{2}\bigg[I_{even}^{\textrm{2mh}}(1)+I^{\textrm{crab}}(1)-I^{\textrm{critter}}(1)+I_{2tri}^{1,3;3,1}-I_{2tri}^{1,3;5,1}-I_{2tri}^{1,3;4,6}+{\rm cyclic}\bigg]\\[1ex]
    &+\frac{\mathcal{C}_{1}+\mathcal{C}^*_{1}}{2\sqrt{2}}\bigg[ I_{odd}^{\textrm{2mh}}(1)-\frac{I_{box;tri}^{4,5,6;1,3}[\epsilon(a,4,5,6,1)]}{(1\cdot 5)}+{\rm cyclic\times2}\bigg]\\[1ex]
    &+\frac{\mathcal{C}_{2}+\mathcal{C}^*_{2}}{2\sqrt{2}}\bigg[ -I_{odd}^{\textrm{2mh}}(2)+\frac{I_{box;tri}^{1,2,3;4,6}[\epsilon(a,1,2,3,6)]}{(2\cdot 6)}+{\rm cyclic\times2}\bigg]\bigg\}
    \end{split}
}
\label{2LoopAnsw}
\end{equation}

\section{Interlude: Infrared regularization using the Higgs mechanism}
\label{sec:higgs}

The two-loop amplitude is infrared divergent and must be regulated in some way.
For inspiration we can look at the four-dimensional sibling of ABJM, $\mathcal{N}=4$ SYM.
In that theory there exists a canonical and self-contained infrared regularization, associated to giving small vacuum expectation values to the scalar fields of the theory \cite{Alday:2009zm}. The fields running in loops then acquire masses through the Higgs mechanism, rendering the loop integrations finite.

Does a similar regularization exist in ABJM theory?
As was shown in the original paper \cite{Aharony:2008ug}, this theory has a moduli space $(\mathbb{C}^4/Z_k)^N$ where $N$
characterizes the SU(N)$\times$ SU(N) gauge group and $k$ is the level.  It is described simply by diagonal vacuum expectation values (vevs) for
the 4 scalar fields, $\l \phi^A\r= \textrm{diag}(v_i^A)$ (with corresponding vevs for the conjugate fields, $\l \bar \phi_A\r = \textrm{diag}((v_i^A)^\dagger)$).
The $Z_k$ identifications will play no role in what follows, although we will be able to see that our formulas are invariant under it.

The first question to address is what is the spectrum of the theory at a given point on the moduli space. While we have not found the general answer to this question in the literature, 
this can easily be answered in perturbation theory in the usual way by studying the linearized action for fluctuations around the vacuum.
(Due to the amount of supersymmetry, it is plausible that the resulting spectrum is valid for all values of the coupling, although this will not be important for us.)
To be safe we have computed the linearized action for both scalar, fermion and gauge field fluctuations, and confirmed that the spectra are related by supersymmetry as required. These computations are reproduced in appendix~(\ref{app:higgs}).
From the linearized action it is then possible to find the poles in the propagators and read off the spectrum.

The result is very simple.  We find that the diagonal fields remain massless, while the off-diagonal fields stretching between $i $ and $j$ acquire the mass squared
\eq
 m_{ij}^2 = (v_i{\cdot}\bar v_i + v_j{\cdot}\bar v_j)^2 -4 v_i{\cdot}\bar v_j v_j{\cdot}\bar v_i.  \label{generalmass}
\eqe
Note that this vanishes when $v_i=v_j$,
as expected.\footnote{More generally, this vanishes whenever the $(Z_k)^N$-invariant combination $v_i\otimes \bar v_i-v_j\otimes \bar v_j$ vanishes.}
Furthermore, when $m_{ij}^2$ is nonzero, the computation in appendix demonstrates that the corresponding
components of the gluon propagator lack a pole at zero momentum.  Thus all modes acquire a mass. This is in contrast with the mass-deformed supersymmetric CSm amplitudes discussed in~\cite{Agarwal:2008pu}. 

\begin{figure}
\begin{center}
\includegraphics{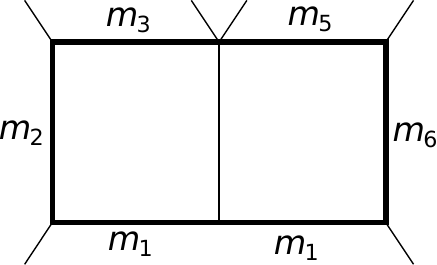}
\caption{Pattern of masses for the Higgsed theory following \cite{Alday:2009zm}.
The loop propagators in the interior of the graph remain massless while those at the boundary, represented in bold, acquire a mass.
External states can be chosen to remain massless or not, depending on whether the $m_i$ are equal or not.}
\label{fig:higgs}
\end{center}
\end{figure}

Following ref.~\cite{Alday:2009zm}, this can be used to regulate planar amplitudes.
The idea is to split SU(N) as SU(M)$\times$ SU(N-M) with $M\ll N$ and turn on vevs only within the smaller SU(M), restricting attention to external states within that SU(M).  Many variants are possible.
For instance if the vev preserves the SU(M) symmetry all external states remain massless.
Or a generic vev can break SU(M) down to U(1)$^M$, rendering the external states massive according to eq.~(\ref{generalmass}).

However, as long as none of the SU(M) vevs vanishes, all outermost propagators in a Feynman diagram will be massive as depicted in fig.~(\ref{fig:higgs}).
This ensures the finiteness of the corresponding integral. (At least for all integrals that have been considered in the literature so far.)

For the purpose of regularization we will restrict to the simplest setup, taking all nonzero vevs to be aligned in the $SU(4)$ directions:
$v_i^A=\delta^A_1 v_i$.  Then the mass formula reduces to
\eq
 m_{ij}^2 =  (m_i -m_j)^2 \qquad \mbox{(aligned vevs)}. \label{massaligned}
\eqe
where $m_i:=|v_i|^2$. We see that the ABJM masses behave exactly like the extra-dimensional coordinates in $\mathcal{N}=4$ SYM discussed in \cite{Alday:2009zm}!

In the remainder of this section, we discuss how to implement this regulator in a simple way
within the embedding formalism.  This will be applied to numerous examples in the next section.

Following the extra-dimensional interpretation of the masses it is natural to enlarge the external five-vectors $y_i$ to six-vectors
\eq
y_i^{(6)} = (x_i,1,x_i^2+m_i^2,m_i)
\eqe
with a inner product defined such that $(i\cdot i)^{(6)}=0$ for vectors of this form.
Then one can verify that $(i\cdot j)^{(6)}=(x_i-x_j)^2+(m_i-m_j)^2$, automatically generating the correct internal masses
provided that the 6-dimensional product is used in propagators. Furthermore, the on-shell constraints are simply
$(i\cdot i{+}1)^{(6)}=0$.  Regarding loop integrations, we set the extra-dimensional component of loop variables to zero, $a=(x,1,x^2,0)$, e.g. the loop variables remain 5-dimensional. Then all propagators come out correctly.

Since only the five-dimensional components of vectors $y_i$ couple to the loop variables, it is immediate that all Feynman parametrization formulas
in section~(\ref{sec:oneloop}) go through unchanged. One must simply continue to use the \emph{five-dimensional} inner product
$(i\cdot j):=(x_i-x_j)^2+m_i^2+m_j^2$ in them.  The \emph{five-dimensional} inner product of the external $y$'s obeys the following identity
\eq
 (i\cdot i{+}1)^2- (i\cdot i)(i{+}1\cdot i{+}1) =  0 \label{onshell}
\eqe
which can be seen to be equivalent to the on-shell relation (\ref{massaligned}) for the external states.

A simple consequence of this procedure is that as long as integrands written in terms of the $y^{(6)}$ are SO(2,3)-covariant,
resulting amplitude will be \emph{invariant} under the modified dual conformal generator
\eq
K^\mu \mathcal{A}_n = 0, \quad\mbox{where}\quad K^\mu = \sum_{i=1}^n \left[ x^2_i \frac{\partial}{\partial x_i^\mu} - 2x_i^\mu x_i{\cdot} \frac{\partial}{\partial x_i}  -2x_i^\mu m_i \frac{\partial}{\partial m_i} - x_i^\mu \right]. \label{modifiedK}
\eqe
This equation is essentially trivial by assumption, and will remain true as long as the integrals are indeed rendered finite by the regularization.

An important question is whether the SO(2,3) symmetry is an actual property of the ABJM integrand even for finite values of the masses.
We expect this to be the case, although we cannot prove it.  The logic is that the dual conformal symmetry SO(2,3) is associated with integrability,
which we do not expect to vanish into thin air just because one moves away from the origin of moduli space.
Indeed, physically, the Higgs branch can be explored by considering amplitudes at the origin of moduli space but with soft scalars added,
as was demonstrated in the context of tree amplitudes in refs.~\cite{Craig:2011ws}.
By exploring such a construction in three dimensions, it might even be possible to establish whether
the dual conformal symmetry of tree amplitudes, hence presumably of loop integrands by unitarity, holds away from the origin of moduli space.\footnote{Dual conformal symmetry of maximal super-Yang-Mills at  finite values of the masses can also be established by considering the symmetry as a property of the higher dimensional parent theory~\cite{CaronHuot:2010rj, Dennen:2010dh}. } 

In the present paper we work only to lowest order in the masses, e.g. we keep only the logarithmic dependence on them. At that level
the SO(2,3) symmetry is more or less tautological as it is the same as the existing dual conformal symmetry. Thus to logarithmic accuracy in the masses
the validity of (\ref{modifiedK}) is already guaranteed by existing results.

Let us elaborate on eq.~(\ref{modifiedK}). A consequence of it together with the on-shell condition (\ref{onshell}) is that the dependence on the individual $m_i$ can be determined simply from consideration of conformal weights.
For instance, suppose an amplitude is known in the case that all internal masses are equal and all external masses vanish.
Then, the most general ``aligned'' case with internal masses $m_i$ (and thus generic external masses) can be obtained (to the same order in the small mass expansion) through the simple substitution
\eq
 \frac{x_{ij}^2}{\mu_\IR^2} \longrightarrow \frac{x_{ij}^2}{m_im_j}. \nonumber
\eqe
The point is that there are no ratios of the masses invariant under (\ref{modifiedK}).
Therefore, with no loss of generality, the Higgs regulator to logarithmic accuracy (and perhaps more generally) can be summarized by the simple rule
\eq
\framebox[14cm][c]{$
\displaystyle{
 y_i \longrightarrow y_i + \mu^2_\IR y_I , \quad \mbox{e.g.}  \quad (\vec x_i,1,x_i^2) \longrightarrow (\vec x_i,1,x_i^2+\mu^2_\IR)
 }$} \label{magicrule}
\eqe
for each external region momenta, where we only need to keep track of the five-dimensional components of the $y_i$, and where 
the massless external momenta remain \emph{undeformed}.
This recipe is the main result of this section.

Alternatively, infrared divergences can be regulated using dimensional regularization. Due to the presence of Levi-Cevita tensors in Chern-Simons theory, dimensional regularization has always been used with a great deal of caution. However, as one can use tensor algebra in three-dimensions to covert the Levi-Cevita tensors into Lorentz invariant scalar dot products and analytically continue to $D=3-2\epsilon$, this is more similar to dimensional reduction regularization, commonly applied to supersymmetric theories.  This regularization scheme has been shown to be gauge invariant up to three-loops for Chern-Simons like theories in ref.~\cite{Chen:1992ee},  and has also been applied to Wilson-loop computations in ref.~\cite{Henn:2010ps,Wiegandt:2011uu} establishing duality with the amplitude result.
We will demonstrate below that the individual dimensionally regulated integrals differ from the mass regulated result functionally. However, when combined into the physical amplitude, the two regulated results agree.

\section{Computation of two-loop integrals} \label{sec:integrals}
In this technical section we describe our computation of the two-loop integrals relevant for the two-loop hexagon.
Although we feel that some of the tricks employed here can find application elsewhere,
the reader not interested in these details can safely skip to the next section. 

Certain integrals, or combinations of integrals, are absolutely convergent and can be computed directly in $D=3$.
Examples are $I^\textrm{2mh}_{odd}$, $I^\textrm{critter}(1)+I^{1,3;4,6}_{2tri}$, or a certain combination of the odd box-triangles described below.
It is very convenient to treat these combinations separately since they can be evaluated without regularization.
Furthermore they automatically give rise to functions of the cross-ratios $u_i$.
On the other hands, IR divergent integrals cannot be avoided.  In this section we will use the Higgs regulator described in the previous section.

We will describe the various steps in our integration method, starting from the steps common to all integrals.

\subsection{A Feynman parametrization trick}
\label{sec:feyn}

There is a particular version of Feynman parameterization which is particularly effective for our calculations.
It is inspired by a formula obtained in \cite{Paulos:2012nu} using intuition from Mellin space techniques,
but can be derived very simply by a judicious change of variable in standard Feynman parameter space
as was demonstrated in ref.~\cite{CaronHuot:2012ab}.

We illustrate it in detail in the case of the double triangle $I_{2tri}^{1,3;3,1}$, the other integrals are entirely similar.
The first step starting from the definition (\ref{doubletris}) is the usual Feynman (Schwinger) trick
\eqa
 I_{2tri}^{1,3;3,1} = \Gamma(3)^2\int_0^\infty \frac{[d^1a_1a_3]}{\textrm{vol(GL(1))}}\frac{[d^1b_1b_3]}{\textrm{vol(GL(1))}}
 \int_{a,b} \frac{(1{\cdot}3)^2}{(a \cdot A)^2(a \cdot b)(b\cdot B)^2}. \label{dt1}
\eqae
where $A=\sum_{i=1,3} a_i y_i$ and $B=\sum_{i=1,3} b_i y_i$.

A word about the notation. The $1/\textrm{vol(GL(1))}$ symbol means to break the projective invariance $(a_i,b_i)\to \alpha (a_i,b_i)$
by inserting any factor which integrates to $1$ on GL(1) orbits. Standard choices include $\delta(\sum a_i-1)$, which give the Feynman parameter measure $dF$ in section (\ref{sec:oneloop}), or $\delta(a_1-1)$ which give rise to Schwinger parameters.
The choice of a gauge-fixing function will play no role in what follows, and for all practical purposes it can be gleefully ignored it until the final step.

The loop integrals over $a,b$ can all be done using only the one-loop integral (\ref{intI0}).
Since the Higgs regulator already renders the integrals finite, we set $D=3$ immediately to obtain (to avoid cluttering the formulas in this section,
we will strip a factor $1/(4\pi)$ for each loop)
\ba
 \Gamma(3)\int_a \frac{1}{(a{\cdot}A)^3} \to \frac{1}{4(\frac12 A\cdot A)^{3/2}}.
\ea
By repeatedly using this formula and its corollary valid for $b^2=0$
\be
 \int_a \frac{1}{(a\cdot A)^2(a\cdot b)} =\Gamma(3)\int_0^\infty df \int_a \frac{1}{(a\cdot (A+fb))^3} = \frac12 \frac{1}{\sqrt{\frac12A\cdot A} (A\cdot b)},
\ee
we derive the following, key formula:
\ba
 \int_{a,b} \frac{1}{(a\cdot A)^2 (a\cdot b)(b\cdot B)^2}
  &=& \frac12\int_b \frac{1}{\sqrt{\frac12A\cdot A} (A\cdot b)(b\cdot B)^2}
 \nl &=& \frac18\int_0^\infty \frac{de}{\sqrt{\frac12 A\cdot A} (\frac12(B+eA)\cdot(B+eA))^{3/2}} \nl
  &=& \int_0^\infty \frac{dc}{4\pi\sqrt c} \int_0^\infty de \frac{1}{ (c \frac12 A\cdot A + \frac12(eA+B)\cdot(eA+B))^2}
  \label{2loopint}.
\ea
The key idea here is the introduction of the new Feynman parameter $c$ in the last step, as done in \cite{CaronHuot:2012ab}.
Although it could be removed immediately, it will prove advantageous to leave it untouched until the final stage. For example,
this will allow us to postpone dealing with square roots until the very end.

Upon substituting (\ref{2loopint}) into (\ref{dt1}), one notes that the variable $e$ is charged under both GL(1) symmetries.
Therefore, it is allowed to gauge-fix one of them by setting $e=1$, which effectively locks the two GL(1) together.
This will always be the case: the variable $e$ is always removable in this way.
Thus we have
\eq
 I_{2tri}^{13;31} = \int_0^\infty \frac{dc}{4\pi\sqrt{c}} \int \frac{[d^3a_1a_3b_1b_3]}{\textrm{vol(GL(1))}}
 \frac{(1\cdot 3)^2}{\big((1+c)\frac12A\cdot A + A\cdot B + \frac12 B\cdot B\big)^2} \label{2tria}.
\eqe
As mentioned, this is similar to the formula for the double-box obtained in \cite{Paulos:2012nu}.

So far all we have done is rewrite the standard Feynman parameter integral in some specific form.
As we will now see, in all cases the variables $a_i, b_i$ can be integrated out rather straightforwardly, and will generate some logarithms or dilogarithms to be
integrated over $c$. The $c$ integration in the final step then poses no particular difficulty.

\subsection{Divergent double-triangles}

Let us see carry out the remainder of this procedure for $I_{2tri}^{1,3;3,1}$ starting from (\ref{2tria}).

Notice that we haven't said anything about the regularization yet.  This is because everything is fully accounted for by the rules (\ref{magicrule}).
According to it, we simply have to take $(i\cdot j)\longrightarrow x_{ij}^2 + 2\mu_\IR^2$.

After evaluating the dot products and doing a simple rescaling of the integration variables, the double-triangle is thus easily seen to depend only on the ratio
$\epsilon:= \frac{\mu^2_\IR}{x_{13}^2}$:
\ba
 I_{2tri}^{13;31} = \int_0^\infty \frac{dc}{4\pi\sqrt{c}} \int \frac{[d^3a_1a_3b_1b_3]}{\textrm{vol(GL(1))}} \frac{1}{
 \big[(a_1{+}b_1)(a_3{+}b_3)+c a_1a_3 + \epsilon \left( (a_1{+}a_3{+}b_1{+}b_3)^2 + c (a_1{+}a_3)^2\right) \big]^2}. \nonumber
\ea
We are interested in the small mass limit $\epsilon\ll 1$.
The only sensitivity to $\epsilon$ comes from the two regions where $a_1,b_1\sim \epsilon$ or $a_3,b_3\sim \epsilon$;
these correspond physically to collinear configurations. However, everywhere else we can ignore $\epsilon$.
Consequently, let us parametrize the variables as
\be
 a_1= 1, \quad b_1= x, \quad a_3= a, \quad  b_3= a y, \quad \frac{[d^3a_1a_3b_1b_3]}{\textrm{vol(GL(1))}} = adadxdy \nonumber
\ee
such that the dangerous regions are $a=0$ and $a=\infty$.
Since the region $a>1$ contributes the same as $a<1$, by symmetry, we need only consider the former and multiply it by 2.
Furthermore, in that region, we can neglect $a$ in the terms proportional to $\epsilon$ since they are only needed when $a\to 0$.
Thus
\ba
 I_{2tri}^{13;31}&=&2\int_0^\infty \frac{dc}{4\pi\sqrt{c}} \int_0^1 ada \int_0^\infty \frac{dxdy}{(a( (1+x)(1+y)+c) + \epsilon ((1+x)^2+c))^2}
 \nl &=&
 \int \frac{dc}{4\pi\sqrt{c}} \int_0^\infty dxdy \frac{\log \frac{(1+x)(1+y)+c}{\epsilon((1+x)^2+c)}-1}{\big((1+x)(1+y)+c\big)^2}
= -\log \frac{4\mu^2_\IR}{x_{13}^2} + \mathcal{O}(\mu_\IR).
\ea
(This was most readily done by evaluating the integrals in the following order: $y$, $x$ and $c$.)

The second type of double-triangle $I_{2tri}^{1,3;3,5}$ is entirely similar.
The general formula (\ref{2loopint}) then gives directly, after a simple rescaling of the variables,
\eqa
 I_{2tri}^{13;35}&=&  \int_0^\infty \frac{dc}{4\pi\sqrt{c}} \int \frac{[d^3a_1a_3b_3b_5]}{\textrm{vol(GL(1))}}
 \nonumber\\ &&\hspace{1cm}\times \frac{1}{
 \big((a_1{+}b_5)(a_3{+}b_3)+a_1b_5+c a_1a_3 + \epsilon' \left( (a_1{+}a_3{+}b_3{+}b_5)^2 + c (a_1{+}a_3)^2\right) \big)^2} \nonumber
\eqae
with $\epsilon'=\frac{\mu^2_\IR x_{15}^2}{x_{13}^2x_{35}^2}$.
The dangerous region is the collinear region $a_1\to0$ and $b_5\to0$, so up to power corrections in $\epsilon'$ we can drop $a_1$ and $b_5$ in the terms
multiplying $\epsilon'$. These can then be easily integrated out, leaving:
\eqa
\!\!\!\!\!  I_{2tri}^{13;35}&=& \int_0^\infty\frac{dc}{4\pi\sqrt{c}} \int_0^\infty db_3 \frac{ \log \frac{(1+b_3)(1+b_3+c)}{(1+b_3)^2+c} -\log\epsilon'}{(1+b_3)(1+b_3+c)}
 =
  1-\frac12\log \frac{4\mu^2_\IR x_{15}^2}{x_{13}^2x_{35}^2} + \mathcal{O}(\mu_\IR).
\eqae

\subsection{A dual conformal integral: $I^\textrm{critter}$}

As our next example, we turn to the integral $I^\textrm{critter}(1)$.
This integral is collinear divergent, but it becomes absolutely convergent after combining it with $I_{2tri}^{1,3;4,6}$ as explained in section (\ref{sec:integrand}).
Therefore, we will only consider the sum
\eqa
\tilde I^\textrm{critter}(1) &:=& I^\textrm{critter}(1) + I_{2tri}^{1,3;4,6}
 \nl &&\hspace{-3cm}= 
 4\int_0^\infty \frac{[d^2a_1a_2a_3]}{\textrm{vol(GL(1))}}\frac{[d^2b_4b_5b_6]}{\textrm{vol(GL(1))}}
 \int_{a,b} \frac{\epsilon(a,1,2,3,*)\epsilon(b,4,5,6,*) + (1\cdot3)(4\cdot6)(a\cdot2)(b\cdot5)}{(a\cdot A)^3(a\cdot b)(b\cdot B)^3}. \label{db1}
\eqae
Note that we have combined the two integrals into a common Feynman parameter integral,
by inserting the inverse propagators $(a.2)(b.5)$ into the numerator of the double-triangle.
This allows us to immediately set the regulating masses to zero, since we are dealing with an absolutely convergent integral.

To apply the formula (\ref{2loopint}), we use the familiar fact that numerators turn into derivatives in Feynman parameter space.
Thus for instance for the first term in (\ref{db1})
\be
4\frac{\epsilon(a,1,2,3,*)\epsilon(b,4,5,6,*)}{(a\cdot A)^3(a\cdot b)(b\cdot B)^3} =\epsilon(\partial_A,1,2,3,*)\epsilon(\partial_B,4,5,6,*)
\frac{1}{(a\cdot A)^2(a\cdot b)(b\cdot B)^2}. \nonumber
\ee
Thus, after setting $e=1$ in (\ref{2loopint}) to remove one of the GL(1) symmetries as done previously, we obtain
\ba
 \tilde I^\textrm{critter}(1) &=& \int_0^\infty \frac{dc}{4\pi\sqrt{c}}
 \int_0^\infty \frac{[d^5a_1a_2a_3 b_4b_5b_6]}{\textrm{vol(GL(1))}}
\bigg(\epsilon(\partial_A,1,2,3,*)\epsilon(\partial_B,4,5,6,*)
\nl && \hspace{0.5cm} {}+(1\cdot3)(4\cdot6)(2\cdot\partial_A)(5\cdot\partial_B)\bigg)
\frac{1}{\big( (c+1)\frac12 A\cdot A + A\cdot B + \frac12 B\cdot B\big)^2}.  \label{bug0}
\ea

To proceed from here, we simply integrate over the variables $a_i, b_i$ one at a time.
This can be done in an essentially automated way using the method described in detail in a four-dimensional context in \cite{CaronHuot:2012ab}.  
The idea is that at each stage the integral can be decomposed into a rational factor which takes the form $dx/(x-x_i)^n$ with $n\geq 1$, times logarithms
or polylogarithms with arguments that are rational functions of $x$. Such integrals can be performed, at the level of the symbol, in a completely automated way.
After this is done, we integrate the symbol and obtain the $c$-integrand as described in \cite{CaronHuot:2012ab}.

We applied this method, doing the integrals in the order $a_2,b_5,a_1,b_6$ and $a_3$, to obtain the symbol of a function to be integrated over $c$.  
After a step of integration by parts in $c$ to remove degree-three components, we obtained the symbol of a degree-two function, which could easily promoted to a function
\eqa
 \tilde I^\textrm{critter}(1) &=& 2\int_0^\infty \frac{dc}{4\pi\sqrt{c}} 
  \frac{\frac{\pi^2}{3}-\Li_2\big(1-u_1(c+1)\big)-\Li_2(1-u_2)-\Li_2(1-u_3)-\log u_2\log u_3}{c+1}  \label{bug2}
\nl &=&
  -\frac12\Li_2(1-u_2)-\frac12\Li_2(1-u_3)-\frac12\log u_2\log u_3 -(\arccos\sqrt{u1})^2 + \frac{\pi^2}{3}. \label{rescritter}
\eqae
Here all non-constant terms come out of the symbol computation, while the $\pi^2/3$ term is a beyond-the-symbol ambiguity. We have fixed it
by an analytic computation at the symmetrical point $u_1=u_2=u_3=1$, where the integral simplifies dramatically. Assuming the principle of maximal transcendentality for this integral, this is the only possible ambiguity. As a cross-check, we have verified that this result agrees with a direct numerical evaluation of eq.~(\ref{bug0}),
to 6 digit numerical accuracy at several random kinematical points with Euclidean kinematics, which we take to confirm our assumptions.%
\footnote{Numerics with this level of accuracy can be easily obtained
starting directly from (\ref{bug0}) and performing the $a_2,b_5$ and $c$ integrals analytically, which are readily done using computer algebra software such as
{\it Mathematica}. The remaining 3-fold numerical integration poses no particular problem.}

\subsection{Another divergent integral: $I^\textrm{2mh}_{even}$}

We now consider a somewhat more nontrivial divergent integral,
\ba
I^\textrm{2mh}_{even}(1) &=&
 \int_0^\infty \frac{dc}{4\pi\sqrt{c}} \int \frac{[d^5a_1a_2a_3b_3b_5b_1]}{\textrm{vol(GL(1))}}
 \frac{\left( \epsilon(\partial_A,1,2,3,*)\epsilon(\partial_B,3,5,1,*)\right)}{\big( (c+1)\frac12 A\cdot A + A\cdot B+ \frac12 B\cdot B\big)^2} \label{2mhfirst}
\ea
where the derivative operators are understood to act on the rational function underneath, to avoid an unnecessary lengthening of the formula.

This integral requires regularization, and as in the rest of this section we use the Higgs regularization described in section (\ref{sec:higgs}).
The procedure has the following precise meaning here. In both the numerator and denominator, we use the shifted five-vectors defined in
eq.~(\ref{magicrule}), so the formula amounts to $(i\cdot j)\longrightarrow(x_i-x_j)^2+2\mu^2_\IR$.

A first observation is that in all divergent regions $b_5\to 0$. Thus we can drop $b_5$ from terms multiplying the mass in the denominator,
which allows us to integrate out $b_5$ explicitly:\footnote{Strictly speaking the numerator derived from eq.~(\ref{2mhfirst}) contains terms proportional to $\mu^2_\textrm{IR}$.
However, due to the special properties of the $\epsilon$-symbol numerators, one can see that these terms only give rise to power-suppressed contributions.
That is, they are never accompanied by compensating $1/\mu^2_\IR$ power infrared divergences which would render them relevant.
We have verified that the same is true also for the integral $I^\textrm{crab}$ considered below.}
\ba
 I^\textrm{2mh}_{even}(1)&=&
 \int_0^\infty \frac{dc}{4\pi\sqrt{c}} \int \frac{[d^4a_1a_2a_3b_1b_3]}{\textrm{vol(GL(1))}} \frac{\big[a_2(2\cdot 5)-2((A+B)\cdot 5)\big] (2\cdot 5)/(1\cdot 3)/((A+B)\cdot 5)^2}
 {\big((1+c)a_1a_3 + a_1b_3+a_3b_1+ b_1b_3 +\frac{\mu^2_\IR}{(1\cdot 3)}X \big)^2}\nonumber
\ea
where $X:=(\sum a+\sum b)^2+c(\sum a)^2$.

To proceed further, we need a small bit of physical intuition about this integral.  It has collinear divergences in the region $a_3,b_3\to 0$ (both loop momenta collinear to $p_1$) and in the region $a_1,b_1\to 0$ (both loop momenta collinear to $p_2$). In addition, there are soft-collinear divergences where these two regions meet.
Thus a reasonable strategy is to subtract something which has the same divergent behavior as $\mu_\IR^2\to0$ but which is simpler to integrate.
A good candidate is
\be
 I^\textrm{2mh}_{even}{}'(1):= I^\textrm{2mh}_{even}(1)\big(X\longrightarrow a_2^2(1+c)\big) \label{defIprime}
\ee
since this remains finite and has identical soft and soft-collinear regions.
But thanks to the simplified denominator, this can be integrated more easily.
Indeed after a shift $a_1+b_1\to b_1$, $a_3+b_3\to b_3$ together with a simple rescaling of the variables,
it can be seen to depend only on a single parameter $\epsilon:=\frac{4\mu_\IR^2 x_{15}^2x_{35}^2}{x_{13}^2x_{35}^4}$:
\eqa
 I^\textrm{2mh}_{even}{}'(1)&=& -\int_0^\infty \frac{dc}{4\pi \sqrt{c}} \int_{\substack{a_1<b_1\\a_3<b_3}}\frac{[d^4a_1a_2a_3b_1b_3]}{\textrm{vol(GL(1))}} \frac{a_2+2b_1+2b_3}{(a_2+b_1+b_3)^2(b_1b_3+a_1a_3 c+\epsilon a_2^2(1+c))^2}
\nl &=& 2-\frac{7\pi^2}{12}-\frac14\log^2 \epsilon + \mathcal{O}(\epsilon). \label{res2mha}
\eqae
From this point we omit further details on the computation of integrals,
as they proceed using the same strategy as in previous examples.
It remains to correct for the error introduced by eq.~(\ref{defIprime}) in the hard collinear regions.
At fixed $a_2,a_1,b_1\sim 1$ one can see that the region $a_3,b_3\to 0$ produces a logarithm whose cutoff depends on $X$.
The error is given by the change in the logarithmic cutoff
\eqa
 I^\textrm{coll}(y)&:=& \int_0^\infty \frac{dc}{4\pi\sqrt{c}} \int_{a_1<b_1}\frac{[d^2a_1a_2b_1]}{\textrm{vol(GL(1))}} \frac{y(a_2y+2b_1)\log \frac{(a_2+b_1)^2+c (a_1+a_2)^2}{a_2^2(1+c)}}{b_1 (b_1+a_1c)(a_2y+b_1)^2}
 \nl &=& 
 \frac{\pi^2}{6}-\Li_2(1-y) \label{Icoll}
\eqae
so that
\eq
 I^\textrm{2mh}_{even}(1) = I^\textrm{2mh}_{even}{}'(1) +  I^\textrm{coll}(x_{25}^2/x_{15}^2)+ I^\textrm{coll}(x_{25}^2/x_{35}^2).
\eqe
This gives the result quoted in appendix~(\ref{app:higgsintegrals}).

\subsection{The integral $I^\textrm{crab}$}

The final divergent integral we have to compute is
\ba
I^\textrm{crab}(1) &=&
 \int_0^\infty \frac{dc}{4\pi\sqrt{c}} \int \frac{[d^5a_1a_2a_3b_5b_6b_1]}{\textrm{vol(GL(1))}}
 \frac{\left( \epsilon(\partial_A,1,2,3,*)\epsilon(\partial_B,5,6,1,*)\right)}{\big( (c+1)\frac12 A\cdot A + A\cdot B+ \frac12 B\cdot B\big)^2}.
\ea
Its evaluation is extremely similar to that in the previous subsection.
The regions which diverge as $\mu^2_\IR\to0$ are the $p_6$-collinear and $p_1$-collinear regions, and their intersection, the soft-collinear
region $A,B\to y_1$.  Therefore, if we denote the $\mu^2_\IR$-containing terms in the denominator by $\mu^2_\IR X$,
we see that we can neglect $a_3$ and $b_5$ in $X$:
\eq
 X=(a_1+a_2+b_6+b_1)^2 +c(a_1+a_2)^2. \nonumber
\eqe
Then we proceed as in the previous example: we replace the integral by the simpler one
\eq
 I^\textrm{crab}{}'(1) := I^\textrm{crab}(1)\big( {X\longrightarrow (a_1+b_1)^2+c a_1^2}\big), \label{subscrab}
\eqe
which we have been able to evaluate as (setting $\epsilon_1:=\frac{4\mu^2_\IR x_{35}^2}{x_{13}^2x_{15}^2}$)
\eq
I^\textrm{crab}{}'(1)=-1-\frac{\pi^2}{12}+\frac12(1+\log u_3)\log \epsilon_1-\frac14\log^2\epsilon_1+\frac12\Li_2(1-1/u_3). \nonumber
\eqe
The error introduced by (\ref{subscrab}) is by construction localized to the hard collinear regions
and turns out to be given by the same eq.~(\ref{Icoll}): $I^\textrm{crab}(1)=I^\textrm{crab}{}'(1)+ I^\textrm{coll}(x_{15}^2/x_{25}^2) +I^\textrm{coll}(x_{13}^2/x_{36}^2)$.
Collecting the terms  gives the result recorded in appendix (\ref{app:higgsintegrals}).

\subsection{Parity odd box-triangles}

As shown in section (\ref{sec:integrand}), parity odd box-triangles appear in the six-point amplitude only in absolutely-convergent combinations
of the form
\eq
\hspace{-0.2cm}
  I^\textrm{odd}_{box;tri}(1) :=  (1\cdot4)(3\cdot 6)\left[\frac{(2\cdot 4)I_{box;tri}^{1,2,3;4,6}[\epsilon(a,1,2,3,6)]}{\epsilon(1,2,3,4,6)}  + \frac{(3\cdot 5)I_{box;tri}^{4,5,6;1,3}[\epsilon(a,4,5,6,1)]}{\epsilon(3,4,5,6,1)}\right]. \label{defIodd}
\eqe
The Feynman parametrization formula (\ref{2loopint}) reads in this case
\ba
I^\textrm{odd}_{box;tri}(1)&=&
 \int_0^\infty \frac{dc}{4\pi\sqrt{c}} \int \frac{[d^5a_1a_2a_3b_4b_5b_6]}{\textrm{vol(GL(1))}}
 \left( \frac{\epsilon(\partial_A,1,2,3,6)}{\epsilon(1,2,3,4,6)} (2\cdot4)(5\cdot\partial_B)
 \right. \nl && \hspace{2cm} \left.
{}+ (2\cdot \partial_A)(3\cdot5)\frac{\epsilon(\partial_B,4,5,6,1)}{\epsilon(4,5,6,1,3)}  \right)
 \frac{(1\cdot4)(3\cdot 6)}{\big( (c+1)\frac12 A^2 + A\cdot B + \frac12 B^2\big)^2}. \nonumber
\ea
where we have combined the two integrals under a common Feynman parameter integral sign.
This may now be evaluated directly without regularization.
One can see that the integrations over $a_2$, $b_5$ do not produce any transcendental functions, which suggests to do them first.
In the process, the $\epsilon$-symbols neatly cancel out:
\ba
I^\textrm{odd}_{box;tri}(1) &=&
\int_0^\infty \frac{dc}{4\pi\sqrt{c}} \int \frac{[d^3a_1a_3b_4b_6]}{\textrm{vol(GL(1))}}
 \left( \frac{a_3(3\cdot5)}{a_1(1\cdot5)+a_3(3\cdot5)}-\frac{b_4(2\cdot4)}{b_4(2\cdot4)+b_6(2\cdot6)}\right)
 \nl && \hspace{2cm} \times
 \frac{(1\cdot4)(3\cdot 6)}{\big((c+1)a_1a_3(1\cdot3) + a_1b_4(1\cdot4)+a_3b_6(3\cdot6)+b_4b_6(4\cdot6)\big)^2}. \nonumber
\ea
The three integrations over $a_i,b_i$ remain elementary, and we obtain a pleasingly simple result
\eq
I^\textrm{odd}_{box;tri}(1)
  = \int_0^\infty \frac{dc}{4\pi\sqrt{c}} \frac{\log \frac{u_2}{u_3} \log(u_1(c{+}1))}{1-u_1(c{+}1)}
  = \frac{\log\frac{u_3}{u_2}\arccos(\sqrt{u_1})}{2\sqrt{u_1(1-u_1)}}.
\eqe

\section{The  six-point two-loop amplitude amplitude of ABJM}\label{section7}
We now construct the final integrated result. We first consider the parity even part, i.e.
terms in eq.~(\ref{2LoopAnsw}) proportional to $\frac{\mathcal{A}^\textrm{tree}}{2}$.  We begin by summing all the divergent integrals,
or more specifically, $\sum_{i=1}^6 (I^\textrm{2mh}_{even}(i)+I^\textrm{crab}(i)+I^{i,i+2;i+2,i}_{2tri}-I^{i,i+2;i+2,i-2}_{2tri})$, using the formulas recorded in appendix (\ref{app:higgsintegrals}). This gives
\eqa
&&
\frac12\sum_{i=1}^6 \left[\log^2\left(\frac{x_{i,i{+}3}^2}{x_{i{+}1,i{+}3}^2}\right) - \log^2\left(\frac{x_{i,i{+}3}^2\mu^2_\IR}{x_{i,i{+}2}^2x_{i{+}1,i{+}3}^2}\right)\right]
+\sum_{i=1}^3 \left[ -\Li_2(1-u_i) +\log u_i \log \frac{x_{i{-}1,i+2}^2}{\mu_\IR^2} \right]
\nl && := BDS_6 - \pi^2.  \label{divintegrals}
\eqae
Pleasingly, we find that all terms of non-uniform transcendentality have canceled in the sum!  Furthermore, the sum
gives nothing but the BDS Ansatz \cite{Bern:1994zx, BDS} in the Higgs regulator!. That is, up to the constant term and the substitution $\mu^2_\IR\to 4\mu^2_\IR$!
The evaluation of the parity even terms will be complete upon adding the dual conformal integrals $-\sum_{i=1}^3 (I^\textrm{critter}(i)+I_{2tri}^{i,i+2;i+3,i-1})$.

As a cross-check on our evaluation of the integrals, we have evaluated the above combination of integrals using dimensional regularization,
which has been successfully implemented in obtaining the two-loop four-point result~\cite{Chen:2011vv, Bianchi:2011dg}.
While we find that the results for individual integrals differ functionally, we find perfect agreement
for the combination just considered, up to an expected scheme-dependent constant.  This constant is given in (\ref{extraconst}).

We next consider the parity odd part, i.e. terms in eq.~(\ref{2LoopAnsw}) proportional to the {\it sum} of one-loop maximal cuts.
The $I_{odd}^\textrm{2mh}$ integrals integrate to zero at order $\mathcal{O}(\epsilon)$, thus we have:
\eqa
\nonumber&&-\frac{\mathcal{C}_{1}+\mathcal{C}^*_{1}}{2\sqrt{2}}\frac{I_{box;tri}^{4,5,6;1,3}[\epsilon(a,4,5,6,1)]}{(1\cdot 5)}
+\frac{\mathcal{C}_{2}+\mathcal{C}^*_{2}}{2\sqrt{2}} \frac{I_{box;tri}^{1,2,3;4,6}[\epsilon(a,1,2,3,4)]}{(2\cdot 4)}+{\rm cyclic\times2}.
\eqae
Using the identity (\ref{A1}) together with the definition (\ref{defIodd}), this can be expressed in terms of the dual conformal invariant finite integral
\eqa
\nonumber&& -\frac{\mathcal{C}_{1}+\mathcal{C}^*_{1}}{2\sqrt{2}} \left[\frac{\epsilon(3,4,5,6,1)}{(1\cdot 4)(3\cdot 6)(1\cdot5)(3\cdot 5)}I^\textrm{odd}_{box;tri}(1)  +{\rm cyclic\times2}\right]\\
&=&\frac{\mathcal{C}_{1}+\mathcal{C}^*_{1}}{2\sqrt{2}} \left[\frac{\epsilon(3,4,5,6,1)}{(1\cdot 4)(3\cdot 6)(1\cdot5)(3\cdot 5)}\frac{\log\frac{u_2}{u_3}\arccos(\sqrt{u_1})}{2\sqrt{u_1(1-u_1)}}  +{\rm cyclic\times2}\right]. \label{oddpart}
\eqae
Due to Yangian invariance, the coefficients of the transcendental functions must be expressible in terms of the leading singularities $LS_1$ and $LS_1^*$ defined in section (\ref{LSdef}).
This can be verified thanks to the remarkable identity (\ref{id3}), together with (\ref{12Rel}):
\eq
\frac{(\mathcal{C}_{1}+\mathcal{C}^*_{1})\epsilon(3,4,5,6,1)}{2\sqrt{2}(1\cdot 4)(3\cdot 6)(1\cdot5)(3\cdot 5)\sqrt{u_1(1-u_1)}} =
\mathcal{A}^\textrm{tree}_{6,\textrm{shifted}}\sgnc\l12\r\sgnc\l45\r
\frac{\big(\l 34\r\l 46\r + \l 35\r\l56\r\big)}{\sqrt{\big(\l 34\r\l 46\r + \l 35\r\l56\r\big)^2}}.\nonumber
\eqe
Thus combining everything, we find the two-loop six-point amplitude of ABJM to be 
\begin{equation}
\boxed{
\begin{split}
  \mathcal{A}_6^{\textrm{2-loop}}&=\left(\frac{N}{ k}\right)^2 \bigg\{\frac{\mathcal{A}_6^{\textrm{tree}}}{2}\bigg[BDS_6+R_6\bigg]
  +\frac{\mathcal{A}^\textrm{tree}_\textrm{6,shifted}} {2}\times\\[1ex]
    &\bigg[ \sgnc (\l12\r) \sgnc(\l45\r)
    \frac{\big(\l 34\r\l 46\r + \l 35\r\l56\r\big)}{\sqrt{\big(\l 34\r\l 46\r + \l 35\r\l56\r\big)^2}}
      \log\frac{u_2}{u_3}\arccos(\sqrt{u_1})+{\rm cyclic\times2}\bigg]
    \bigg\}
    \end{split}
} \label{result1}
\end{equation}
where the ``remainder'' function $R_6$ is given as 
$$R_6=-2\pi^2+\sum_{i=1}^3\left[\Li_2(1-u_i)+\frac12\log u_i\log u_{i{+}1} +(\arccos\sqrt{u_i})^2\right].$$
We like to stress that the remainder function, up to an additive constant, is given entirely by the dual-conformal finite integral $\tilde I^\textrm{critter}(i)$ for the parity even structure (and by $I^\textrm{odd}_{box;tri}(i)$ for the parity-odd structure).
This is in contrast with $\mathcal{N}=4$ SYM, where the remainder function is mixed with BDS and spread across a number of divergent integrals. 

The part proportional to $\mathcal{A}^\textrm{tree}_{6,\textrm{shifted}}$ can be written in a more compact way if we assume certain restrictions on
the kinematics.  We will assume so-called Euclidean kinematics, e.g. all non-vanishing invariants $(i\cdot j)$ are spacelike.
(This is a nonempty region even for real Minkowski momenta.)
In that case, it is correct to naively rewrite the original expression in terms of angle-brackets:
\eq
\frac{\arccos(\sqrt{u_1})}{\sqrt{u_1(1-u_1)}} := \frac{1}{2i} \frac{\log\left(\frac{\sqrt{u_1}+i \sqrt{1-u_1}}{\sqrt{u_1}-i\sqrt{1-u_1}}\right)}{\sqrt{u_1(1-u_1)}}
\longrightarrow \frac{(1\cdot 4)(3\cdot 6)\log \chi_1}{2i \l 12\r\l45\r\big(\l 34\r\l46\r+\l35\r\l46\r\big)} \nonumber
\eqe
where
\eq
 \chi_1:=\frac{\l12\r\l45\r + i(\l34\r\l46\r+\l35\r\l56\r)}{\l12\r\l45\r-i (\l34\r\l46\r+\l35\r\l56\r)}. \label{defChi}
\eqe
Indeed, in that region, $u_1>0$ and the first expression is always real and positive.  This is also the case for the second expression, as can be seen
from the fact that $\l12\r$ and $\l 45\r$ are real, while $(\l34\r\l46\r+\l35\r\l56\r)$ is either real or smaller in magnitude than $\l12\r\l45\r$.
Note that, defining cross-ratios $\chi_2$ ($\chi_3$) from cyclic shifts by minus 2 (plus 2) of this expression, it can be shown that $\chi_1\chi_2\chi_3=1$.

This allows us, in these kinematics, to simplify the answer to
\begin{equation}
\boxed{
\begin{split}
  \mathcal{A}_6^{\textrm{2-loop}}&=\left(\frac{N}{ k}\right)^2 \bigg\{\frac{\mathcal{A}_6^{\textrm{tree}}}{2}\bigg[BDS_6+R_6\bigg]
  +\frac{\mathcal{A}^\textrm{tree}_\textrm{6,shifted}} {4i}
    \bigg[\log\frac{u_2}{u_3}\log\chi_1+{\rm cyclic\times2}\bigg]  \bigg\}\,.
    \end{split}} \label{result2}
\end{equation}
While strictly derived from eq.~(\ref{result1}) in Euclidean kinematics, we expect this expression
to be valid in other kinematic regions for a suitable analytic continuation of the variables $\chi_i$. Note that as $\chi_i\rightarrow\chi_i^{-1}$ under the $Z_2$ little group transformation of any external leg, the little group weight of $\log\chi_1$ is exactly what is needed to compensate the little group mismatch of $\mathcal{A}^\textrm{tree}_\textrm{6,shifted}$. 

\section{Conclusions}\label{section8}
In this paper, we construct the two-loop six-point amplitude of ABJM theory. The result can be separated into a two-loop correction proportional to the tree amplitude, and a correction proportional to the shifted tree amplitude, which are distinct Yangian invariants. The correction proportional to the first is infrared divergent and we use mass regularization. The result shows that the infrared divergence is identical to that of $\mathcal{N}=4$ super Yang-Mills and is thus completely captured by the BDS result. This establishes that the dual conformal anomaly equation is identical between the one-loop SYM$_4$ and two-loop ABJM, which was first observed at four-points and we conjecture will persists to all points. The correction multiplying the shifted tree amplitude is completely finite.

As a comparison, we also computed the divergent integrals using dimensional reduction. We find that the individual integrals give different functional answer between the two regularization schemes. However, when combined into the amplitude, they give the same result up to a physically expected constant.

We find in addition to the BDS result a nonzero (dual-conformal invariant) remainder function.
This implies that the six-point ABJM amplitude cannot be dual to a bosonic Wilson-loop, which only captures the BDS part~\cite{Wiegandt:2011uu}%
\footnote{Note that the vanishing of 1-loop Wilson loops was obtained numerically in \cite{Henn:2010ps}.  Given the subtle
analytic properties discussed in section (\ref{sec:analysis}), it could be worthwhile to supplement this by an analytic computation.}.
This does not rule out a possible duality with a suitable supersymmetric Wilson loop, however.
The reason is that, if SYM${}_4$ is to be of any guidance \cite{Mason:2010yk}, the correct Wilson loop dual for amplitudes with $n\geq 6$
particles should reproduce, at lowest order in the coupling, the $n$-point tree amplitude\footnote{At least up to
$\delta^3(P)\delta^6(Q)$ and a purely bosonic factor, akin to the Parke-Taylor denominator in SYM${}_4$.}.
Since no candidate Wilson loop with this property, or even just the correct quantum numbers,
are presently available in the literature, we find it hard to say anything conclusive about the duality.
Our results demonstrate that the dual conformal symmetry persists at the quantum level up to an anomaly which is identical to that
of a Wilson loop.  We interpret this as strong evidence for the existence of a dual Wilson loop which remains to be constructed.

We list a number of open questions for future work.
A first one concerns the status of the dual conformal symmetry away from the origin of moduli space, e.g. in the Higgsed theory.
As demonstrated in section~(\ref{sec:higgs}), to lowest order in the masses (logarithmic accuracy), the Higgsed theory enjoys an exact dual conformal symmetry
under which the masses transform in a nontrivial way.
It is not clear whether this symmetry extends all the way into the moduli space; for one thing, the origin of the symmetry is mysterious and the original
string theory argument in \cite{Alday:2009zm} does not apply in ABJM due to difficulties with the T-duality.
As discussed in the main text, a key step here would be to settle this question for the tree amplitudes.

We note that a 3-loop computation of the 4-point and 6-point amplitude in ABJM would probably be feasible with the same techniques,
although a more sustained effort would be required.
For instance, we expect only degree-3 transcendental functions in the result.
Furthermore, the only divergences should be double-logarithms multiplying the 1-loop amplitude.
Given the absence of overlapping divergences, the integration technology developed in section (\ref{sec:integrals}) might thus plausibly be sufficient. 

An interesting property of our remainder function $R_6$ is that it does \emph{not} vanish in collinear limits, contrary to the case in SYM${}_4$.
In fact, it even diverges logarithmically in the `simple' collinear limit (six point goes to five),
this even though the five-point amplitude is zero.
This does not violate any physical principle, since the $A^\textrm{tree}_6$ and $A^\textrm{tree}_{6,\textrm{shifted}}$ prefactors do not have any pole in this limit.
In the absence of a pole, there is no need for the amplitude to factorize into a product of lower-point amplitudes.
In other words, the \emph{leading} term in the collinear limit in ABJM is similar to subleading, \emph{power-suppressed} terms in the collinear limits in $D=4$.
The factorization theory for these terms is more complicated, and in fact it has only been worked out recently in the dual Wilson loop language \cite{Alday:2010ku}.
It would be very interesting to work out the general structure of this limit, using field theory arguments, as this should place strong constraints on the amplitudes.
In subsection (\ref{sec:analysis}), for instance, we have conjectured from analyticity of the scattering amplitudes that a certain discontinuity of the amplitude should vanish in the collinear limit, but we have no idea how this could be established.

Also interesting are the `double-collinear' limit (six point goes to four), or factorization limits ($p_{123}^2$ goes to zero, but momenta $p_{1,2,3}$ do not become collinear).
Since it was not clear to the authors what kind of field theory predictions are available for these limits, we did not discussed them on our 6-point result.
However, it is possible that this could shed further light on our result itself, for instance by giving a physical interpretation for
the relative signs between different terms.  These limits may also yield some interesting constraints on the higher-point amplitudes.

Another interesting direction for future work concerns the rest of the Yangian algebra at loop level.
As one easily sees from \cite{Huang:2010qy}, the Yangian algebra in ABJM is generated by the bosonic dual conformal symmetry
together with the (ordinary) superconformal symmetry.  Since the former is presently conjectured to become anomaly-free to all loops after
dividing by the BDS Ansatz, the crux is the superconformal anomaly.  By analogy with SYM${}_4$,
the properly understood symmetry at the quantum level should uniquely determine the amplitudes, providing for an efficient way to compute them.
Our two-loop result (\ref{result2}) should thus provide an important data point to understand the quantum symmetries of ABJM,
perhaps combining the 1-loop ABJM analysis in \cite{Bargheer:2012cp} with the all-loop SYM${}_4$ analysis in \cite{CaronHuot:2011kk}.

\section{Aknowledgement}
We would like to thank Johannes Henn for many enlightening discussions. 
Y-t would like to thank N.~Arkani-Hamed for invitation as visiting member at the Institute for 
Advanced Study at Princeton, where this work was initiated.
SCH gratefully acknowledges support from the Marvin L.~Goldberger Membership and from
the National Science Foundation under grant PHY-0969448.
This work was supported in part by the US Department of Energy under contract DEÐFG03Ð 91ER40662. 
\appendix
\section{Identities \label{IdApp}}
In this appendix, we aim to prove a series of identities used in the text. First consider the following identity:
\eq
\frac{\mathcal{C}_{1}+\mathcal{C}^*_{1}}{\mathcal{C}_{2}+\mathcal{C}^*_{2}} = -\frac{\epsilon(6,1,2,3,4)(3\cdot5)(5\cdot1)}{\epsilon(3,4,5,6,1)(6\cdot2)(2\cdot4)}\,.
\label{A1}
\eqe
The strategy is to express the five-dimensional $\epsilon$ symbol in terms of angle brackets.
To do so, we start from the definition of the $\epsilon$-symbol as a determinant
and use the manifest translation invariance of the formula to set $x_2=0$.
In doing so, we must remember to normalize the determinant such that $\epsilon(i,j,k,l,m)^2$ agrees with the Gram determinant formula (\ref{note1}),
since this is the convention used in the main text; this requires an extra factor of $2i\sqrt{2}$. Thus
\def\vp{\vec{p}}
\eq
 \epsilon(6,1,2,3,4) := 2i\sqrt{2}\det (y_6,y_1,y_2,y_3,y_4 ) =
 2i\sqrt{2}\det \left(\begin{array}{c@{\hspace{0.3cm}}c@{\hspace{0.3cm}}c@{\hspace{0.3cm}}c@{\hspace{0.3cm}}c}
 -\vp_6{-}\vp_1 & -\vp_1 & 0 & \vp_2 & \vp_2{+}\vp_3 \\
  1 & 1 & 1 & 1 & 1 \\
  -\l 61\r^2 & 0 & 0 & 0 & -\l23\r^2 \end{array}\right). \nonumber
\eqe
where the first three rows are real in Minkowski signature.
This determinant can now be evaluated in terms of three-dimensional ones, which in turn give two-brackets:
$\det(\vp_i,\vp_j,\vp_k):=\frac 12 \l ij\r\l jk\r\l ik\r$.  This way we obtain
\eq
 \epsilon(6,1,2,3,4) = i\sqrt{2}\l 61\r \l 12\r\l23\r \big(\l 31\r\l 16\r+ \l32\r\l26\r\big).
\label{id1}
\eqe
Performing a similar computation for $\epsilon(3,4,5,6,1)$ and using that $(3\cdot 5)=-\l34\r^2$ etc., we thus find
\eq
\frac{\epsilon(6,1,2,3,4)(3\cdot5)(5\cdot1)}{\epsilon(3,4,5,6,1)(6\cdot2)(2\cdot4)} = \frac{\l12\r\l34\r\l56\r}{\l 23\r\l45\r\l61\r}
 \left(\frac{\l 31\r\l 16\r+ \l32\r\l26\r}{\l 34\r\l 46\r+ \l35\r\l56\r}\right).  
\eqe
Using momentum conservation, the parenthesis can be shown to equal $-1$, proving the desired formula using (\ref{12Rel}).

Another remarkable algebraic identity is
\eq
  x_{14}^2x_{36}^2-x_{13}^2x_{46}^2 = (\l 34\r\l 46\r + \l 35\r\l56\r)^2
\label{id2}
\eqe
which one might call a Dirac matrix trace identity,
and follows from squaring eq.~(\ref{id1}) and using that the square should give the Gram determinant. Using this identity we have that
\eq
\frac{\l12\r\l34\r\l56\r\epsilon(3,4,5,6,1)}{(1\cdot 4)(3\cdot 6)(1\cdot5)(3\cdot 5)\sqrt{u_1(1-u_1)}} =
\frac{i\sqrt{2}\l12\r\l45\r \big(\l 34\r\l 46\r + \l 35\r\l56\r\big)}{\sqrt{\l12\r^2\l 45\r^2\big(\l 34\r\l 46\r + \l 35\r\l56\r\big)^2}}
\label{id3}
\eqe
which was used around eq.~(\ref{oddpart}).
\section{ABJM theory on the Higgs branch}
\label{app:higgs}

The action of ABJM takes the form (see for instance \cite{Minahan:2008hf} for an explicit component form):
\eq
\LL = \frac{k}{4\pi} \big( \LL_\textrm{kin} + \LL_{4} + \LL_{6}\big).
\eqe
To describe the spectrum of the theory on the Higgs branch, we begin by describing the fermion mass matrix.
The interactions of the fermions can be written, following \cite{Minahan:2008hf} but
as can also be verified directly by comparing against various components of the four-point amplitude (\ref{fourpoint}),
\eqa
  \LL_{4}&=& \left(\textrm{Tr}[\psi_A \bar \psi^B \phi^C \bar\phi_D]-\textrm{Tr}[\psi_A \bar \phi_D \phi^C \bar\psi^B]\right) (2\delta^A_C\delta_B^D-\delta^A_B \delta_C^D)
  \nl && + \epsilon_{ABCD}\textrm{Tr}[\phi^A \bar\psi^B\phi^C \bar\psi^D]+\epsilon^{ABCD}\textrm{Tr}[\psi_A \bar\phi_B\psi_C \bar\phi_D] .  \label{interactionf}
\eqae
As already mentioned in the main text, the moduli space $(\mathbb{C}^4/Z_k)^{N}$ of this theory is characterized by diagonal vacuum expectation values for the scalar fields. Let us denote the fields above (below) the diagonal with a plus (minus) superscript, so that $(\psi_A^\pm)^\dagger=\bar\psi^{A\mp}$. As one can easily see from the action (\ref{interactionf}), the diagonal fermions remain massless while $\psi_B^+$ and $\bar\psi^{B+}$ mix with each other.
Upon inserting a diagonal vev for the scalars, the mass term thus takes the form
$(\bar\psi^{A-},\psi_A^-)M_f(\psi_B^+,\bar\psi^{B+})^T$ where $M_f$ is the $8\times8$ Hermitian matrix
\eq
 M_f = \left(\begin{array}{cc} 2(x\bar x-y\bar y)^B_A - \delta^B_A (x{\cdot}\bar x - y{\cdot}\bar y) & 2\epsilon_{ABCD} x^C y^D \\
 2\epsilon^{ABCD}\bar y_C \bar x_D & 2(y\bar y-x\bar x)_B^A - \delta_B^A (y{\cdot}\bar y - x{\cdot}\bar x)\end{array}\right). \nonumber
\eqe
In this appendix, $(x,y)^A:=(v_i,v_j)^A$ will denote the diagonal vevs coupled to off-diagonal components under consideration.
As one can verify $M_f^2=\idmatrix_8 m^2$ with $$m^2= (x{\cdot}\bar x + y{\cdot}\bar y)^2 - 4 x{\cdot}\bar y y{\cdot}\bar x,$$
showing that all 8 off-diagonal fermions acquire the same mass. 

The scalar potential was described in detail in ref.~\cite{Aharony:2008ug},
\eq
 \LL_{6} = \Tr[\phi^A \bar\phi_{[A} \phi^B \bar\phi_{C]}\phi^C\bar\phi_B] - \frac1{3} \Tr[ \phi^A \bar\phi_{[A}\phi^B \bar\phi_{B} \phi^C \bar\phi_{C]}] \nonumber
\eqe
where here the square bracket means antisymmetrization in the indices. Again one can see that the diagonal fluctuations remain massless
while off-diagonal ones $\delta\phi^{A+}$ and $\delta\bar\phi_A^+$ mix with each other.
It follows that the mass term takes the form $(\delta\bar\phi_A^-,\delta\phi^{A-}) M^2_s (\delta\phi^{B+},\delta\bar\phi_B^+)^T$, and a computation gives
the $8\times 8$ Hermitian matrix as
\eq
 M^2_s = m^2 \left(\idmatrix_8 - \mathbb{P}_8 \right)
\eqe
where $\mathbb{P}_8=\left(\begin{array}{c@{\hspace{0.3cm}}c} x^T&y^T\\ -\bar y^T&-\bar x^T\end{array}\right){\cdot}
\left(\begin{array}{cc} x{\cdot}\bar x + y{\cdot}\bar y & -2x{\cdot}\bar y\\ -2y{\cdot}\bar x \bar y&x{\cdot}\bar x + y{\cdot}\bar y\end{array}\right)
{\cdot}
\left(\begin{array}{c@{\hspace{0.4cm}}c} \bar x&-y\\ \bar y&-x\end{array}\right)/m^2$ is an orthogonal projector onto the two would-be Goldstone bosons
$(\delta\phi^+,\delta\bar\phi^+)\sim (x,-\bar y)$ and $(\delta\phi^+,\delta\bar\phi^+)\sim(y,-\bar x)$.
We conclude that six of the eight scalars acquire the same mass squared as the fermions,
while the remaining two acquire no mass, although they are soon to be ``eaten'' by the gauge fields through the Higgs mechanism.

Finally, we consider the gauge fields, which acquire mass terms through the scalar kinetic term $\sim \Tr D_\mu \phi D^\mu \bar\phi$.
First we discuss the off-diagonal components.
As one can see again the fields $A_{1,2}^+$ from the two gauge groups mix with each other,
so the mass term is characterized by a $2\times2$ Hermitian matrix $(A_1^-,A_2^-)M_g(A_1^+,A_2^+)$.
However, in this case the kinetic term is also characterized by a nontrivial matrix $d(A_1^-,A_2^-) \wedge K_g (A_1^+,A_2^+)$.
These two matrices are
\eq
 K_g= \left(\begin{array}{cc} 1&0\\0&-1\end{array}\right) \quad\mbox{and}\quad 
 M_g= \left(\begin{array}{c@{\hspace{0.3cm}}c} x{\cdot}\bar x + y{\cdot}\bar y & -2x{\cdot}\bar y\\ -2y{\cdot}\bar x & x{\cdot}\bar x + y{\cdot}\bar y \end{array}\right).\nonumber
\eqe
Fortunately, to obtain the propagator it is not necessary to diagonalize these two matrices simultaneously
--- as pointed out in \cite{Mukhi:2011jp} this may not even be possible in general.
In the present case, one can verify that $(K_g M_g)^2 = m^2\idmatrix_2 $ and this suffices in order to write down the propagator in a simple way.
To see this, let us first add a gauge-fixing term to the action
$(\partial_\mu A^{\mu-} - \xi v{\cdot}(\delta\phi^-))K_g (\partial_\mu A^{\mu+} - \xi v{\cdot}(\delta\phi^+))/\xi$, designed to remove the mixing between
the gauge bosons and the scalar fields, where $\xi$ is some arbitrary scale.
Then the two unphysical scalars acquire masses squared $\sim \xi m$, and a short computation gives the gluon propagator as
\eq
 \l (A_1^{\mu-},A_2^{\mu-})(p) \left(\begin{array}{c} A_1^{\nu+} \\ A_2^{\nu+}\end{array}\right) \r \propto
 \frac{\epsilon^{\mu\nu\sigma}p^\sigma K_g + \delta^{\mu\nu} K_gM_gK_g}{p^2+(K_gM_g)^2}
 + p^\mu p^\nu\frac{\xi K_g +K_gM_gK_g \frac{\xi^2-p^2}{p^2+(K_gM_g)^2}}{p^4 +\xi^2 (K_gM_g)^2}. \nonumber
\eqe
In particular, this formula shows that there are no singularities at zero momentum provided $m^2\neq 0$, as required in the main text.

Finally, we discuss the diagonal gauge fields. Since only the combination $(A_1-A_2)$ receives a mass term
in this case, we have that
$M_g\propto (1,-1)\otimes (1,-1)^T$ which is effectively nilpotent: $(K_gM_g)^2=0$.
As the above propagator shows,  even though the mass matrix is nonzero,
no massive states appear in the spectrum (as required by supersymmetry).
This situation has been discussed in detail 
in \cite{Mukhi:2011jp}.\footnote{
Note that in refs.~\cite{Mukhi:2011jp} it was further shown that the field $(A_1-A_2)$ can be integrated out in a systematic expansion in $1/m$,
yielding a Yang-Mills term $k F^2/m$ for the remaining gauge field plus other terms.
But since for us $m\sim |v_i|^2$ is an infrared scale, not an ultraviolet scale, such a (in any case not strictly necessary) procedure would be inappropriate in our context.}

\section{Integrals using the mass regularization}
\label{app:higgsintegrals}

Here we summarize the results obtained in section~(\ref{sec:integrals}) for the integrals defined in eqs.~(\ref{defI6}), multiplied by $16\pi^2$,
evaluated using a small internal mass to regulate infrared divergences as defined in section~(\ref{sec:higgs}).
\eqa
I^\textrm{crab}(1) &=&-1+\frac{\pi^2}{4} +\frac12(1+\log u_3)\log \frac{4\mu^2_\IR x_{35}^2}{x_{13}^2x_{15}^2} -\frac14\log^2\frac{4\mu^2_\IR x_{35}^2}{x_{13}^2x_{15}^2} 
 \nonumber\\
&&-\Li_2(1-x_{13}^2/x_{36}^2)-\Li_2(1-x_{15}^2/x_{25}^2)+\frac12\Li_2(1-1/u_3),
 \nonumber\\
 I^\textrm{2mh}_{even}(1) &=& 2-\frac{\pi^2}{4}-\frac14\log^2 \frac{4\mu^2_\IR x_{15}^2x_{35}^2}{x_{13}^2x_{25}^4}-\Li_2(1-x_{25}^2/x_{15}^2)-\Li_2(1-x_{25}^2/x_{35}^2),
\nonumber\\
I_{2tri}^{13;13} &=& -\log \frac{4\mu^2_\IR}{x_{13}^2},
\nonumber\\
I_{2tri}^{13;35} &=& 1-\frac12\log \frac{4\mu^2_\IR x_{15}^2}{x_{13}^2x_{35}^2},
 \nonumber\\
I^\textrm{2mh}_{odd}(1) &=& 0. \label{reshiggs}
\eqae
In addition, we have the following two absolutely-convergent integrals:
\eq
I^\textrm{critter}(1)+I_{2tri}^{13;46}=-\frac12\Li_2(1-u_2)-\frac12\Li_2(1-u_3)-\frac12\log u_2\log u_3 -(\arccos(\sqrt{u_1}))^2+\frac{\pi^2}{3}
\eqe
and (see eq.~(\ref{defIodd}) for the definition)
\eq
I^\textrm{odd}_{box;tri}(1) = \frac{\log\frac{u_3}{u_2}\arccos(\sqrt{u1})}{2\sqrt{u1(1-u1)}}.
\eqe

For completeness, we find for the four-point double-box using the same regularization:
\eq
 I_{2box}^{1,2,3;3,4,1}[\epsilon(a,1,2,3,*)\epsilon(b,3,4,1,*)] = \frac{\pi^2}{3} +\log \frac{4\mu^2_\IR}{x_{13}^2} -\log\frac{4\mu^2_\IR}{x_{13}^2}\log\frac{4\mu^2_\IR}{x_{24}^2}.
\eqe

\section{Integrals using dimensional regularization\label{DimReg}}
In this appendix, we present the integrated result of infrared divergent integrals using dimensional regularization. Here all integrals are again multiplied by $16\pi^2$. After obtaining the integrals in terms of Feynman parameters,  we integrate by converting the integrand into Mellin-Barnes representation, and implement the Mathemtica package MB.m~\cite{MB} to obtain the result up to $\mathcal{O}(\epsilon)$. The result is expressed in terms of zero-, one- and two-dimensional integrals in Mellin space. The one and two-dimensional integrals are analytically evaluated by performing sum over residues. That such sum can be carried out analytically, is simply due to the fact that the two-loop amplitude should be of transcendental two functions.

The two mass hard integral gives: 
\eqa
\nonumber&&I^{\textrm{2mh}}_{even}(1)=\int \frac{\epsilon(a,1,2,3,*)\epsilon(b,3,5,1*)}{(a\cdot 1)(a\cdot 2)(a\cdot3)(a\cdot b)(b\cdot3)(b\cdot5)(b\cdot1)}\\
\nonumber&=&\frac{e^{-\gamma\,2\epsilon}}{(4\pi)^{-2\epsilon}}\bigg[\frac{(2)^{2 \epsilon}(3 (x_{13}^2)^{-2 \epsilon}-(x_{15}^2)^{-2 \epsilon} - 
    (x_{35}^2)^{-2 \epsilon} + 2(x_{25}^2)^{-2\epsilon})}{
   16 \epsilon^2}-\frac{(2)^{-2 \epsilon}}{
   16 \epsilon^2}\left( \frac{(x_{13}^2) (x_{15}^2)(x_{35}^2)}{(x_{25}^2)^{2}}\right)^{-2 \epsilon}\bigg]\\
 \nonumber&& +\frac{1}{8}\bigg[\log^2(x^2_{13}/x^2_{15})-\log^2(x^2_{13}/x^2_{35})+2\log^2(x^2_{13}/x^2_{25}) \\
  \nonumber&& +2\log^2(x^2_{15}/x^2_{25})-2\log^2(x^2_{35}/x^2_{25})+3\log^2(x^2_{15}/x^2_{35})\bigg] \\
\nonumber&&+\pi\arcsin(\sqrt{x^2_{15}/x^2_{25}}) -\arcsin^2(\sqrt{x^2_{15}/x^2_{25}})+\pi\arcsin(\sqrt{x^{2}_{35}/x^2_{25}})-\arcsin^2(\sqrt{x^{2}_{35}/x^2_{25}})\\
 && -\frac{1}{2}Li_2(1 - x^2_{15}/x^2_{25})-\frac{1}{2}Li_2(1 -x^{2}_{35}/x^2_{25})+\frac{11\pi^2}{24}+\frac{\log^22}{4}-2-2\pi^2*a 
 \eqae
where $a\approx0.3594267177020808$. The crab integral gives:
\eqa
\nonumber &&I^{\textrm{crab}}_4(1)=\int \frac{\epsilon(a,1,2,3,*)\epsilon(b,5,6,1,*)}{(a\cdot 1)(a\cdot 2)(a\cdot3)(a\cdot b)(b\cdot5)(b\cdot6)(b\cdot1)}
\\
\nonumber&=&\frac{e^{-\gamma\,2\epsilon}}{(4\pi)^{-2\epsilon}}\bigg[\frac{(x_{13}^2)^{-2 \epsilon}+(x_{15}^2)^{-2 \epsilon}+(x_{26}^2)^{-2\epsilon}-
    (x_{25}^2)^{-2 \epsilon}-(x_{36}^2)^{-2\epsilon}}{8\epsilon^2}+\frac{(x_{35}^2)^{-2\epsilon}-(x_{13}^2)^{-2 \epsilon}-(x_{15}^2)^{-2 \epsilon}}{4\epsilon} \bigg]\\
\nonumber&&-\frac{1}{4}\bigg[\log^2( x^2_{13}/x^2_{15})-\log^2( x^2_{13}/x^2_{25})-\log^2( x^2_{15}/x^2_{36})-\log^2( x^2_{15}/x^2_{25})-\log^2( x^2_{13}/x^2_{36})\bigg]\\
\nonumber&&-\frac{1}{4}\bigg[\log^2( x^2_{26}/x^2_{13})+\log^2( x^2_{26}/x^2_{15})+\log^2( x^2_{25}/x^2_{35})+\log^2( x^2_{36}/x^2_{35})-\log^2( x^2_{35}/x^2_{26})\bigg]\\
\nonumber&&-\pi\arcsin(\sqrt{x^2_{15}/x^2_{25}})+\arcsin^2(\sqrt{x^2_{15}/x^2_{25}})-\arcsin(\sqrt{x^2_{13}/x^2_{36}})+\arcsin^2(\sqrt{x^2_{13}/x^2_{36}})\\
&&+\frac{1}{2}\bigg[\frac{1}{2}\log^2\left( u_3\right)+Li_2(1 - x^2_{15}/x^2_{25})+Li_2(1 -x^2_{13}/x^2_{36})+Li_2\left(1 - u_3\right)\bigg]+\frac{\pi^2}{48}+\frac{1}{2}
\eqae
Notice the presence of $\arcsin$ functions with non-conformal cross-ratios as arguments. Such function did not appear in the mass regulated result and marks a stark distinction between the two regularizations. The ArcSin functions always come in the combination 
$$\arcsin(\sqrt{m})-\arcsin^2(\sqrt{m})/\pi\,.$$ 
This particular combination is necessary for the integral to remain real.

For completeness, we list the double triangle result:
\eqa
\nonumber I_\textrm{2tri}^{1,3;3,1}&=&\int\frac{(1\cdot3)^2}{(a\cdot1)(a\cdot3)(a\cdot b)(b\cdot1)(b\cdot3)}
=-\left(\frac{e^{\gamma}x_{13}^2}{4\pi}\right)^{-2 \epsilon}\frac{1}{2\epsilon}-1\,,\\
\nonumber I_\textrm{2tri}^{1,3;5,1}&=&\int\frac{(1\cdot3)(1\cdot5)}{(a\cdot1)(a\cdot3)(a\cdot b)(b\cdot5)(b\cdot1)}
=-\left(\frac{e^{\gamma}x_{13}^2x^2_{15}}{4\pi\,x^2_{35}}\right)^{-2 \epsilon}\frac{1}{4\epsilon}+\frac{1}{2}\,.
\eqae

While the integrated results appears to be regularization scheme dependent (compare with eqs.~(\ref{reshiggs})), when combined into amplitudes they give identical result up to additive constants. In particular, the $\arcsin$ functions completely cancel. Considering the sum of infrared divergent integrals in eq.~(\ref{divintegrals}) one obtains:
\eqa
\nonumber&&\sum^6_{i=1}\;\;I_{even}^{\textrm{2mh}}(i)+I^{\textrm{crab}}(i)+I_{2tri}^{i,i+2;i+2,i}-I_{2tri}^{i,i+2;i+4,i}\\
\nonumber&=&\bigg[\sum^6_{i=1}\bigg(-\frac{e^{-\gamma2\epsilon}}{(8\pi)^{-2\epsilon}}\frac{(x^2_{i,i+2})^{-2\epsilon}}{(2\epsilon)^2}-\log\frac{x^2_{ii+2}}{x^2_{ii+3}}\log\frac{x^2_{i+1i+3}}{x^2_{ii+3}}+\frac{1}{4}\log^2\frac{x_{ii+3}}{x_{i+1i+4}}\\
\nonumber&&-\frac{1}{2}\Li_2\left(1 - \frac{x^2_{ii+4}x^2_{i+1i+3}}{x^2_{ii+3}x^2_{i+1i+4}}\right)\bigg)-\pi^2(\frac{23}{8}-12*a)\bigg]\\
&=&BDS_6(\epsilon\rightarrow2\epsilon)-\pi^2(\frac{31}{8}-12*a) \label{extraconst}
\eqae
where $BDS_6$ is the one-loop six-point MHV amplitude of $\mathcal{N}=4$ sYM~\cite{Bern:1994zx} with $\epsilon$ replaced by $2\epsilon$,  reflecting the two-loop nature of the result. Thus the dimensionally regulated infrared divergent integrals combine to give the BDS answer, just as in the mass regulated result.

\end{document}